\documentclass[referee]{aa}
\usepackage{graphicx}
\usepackage{epsfig}

\begin{document}
   \title{Slow evolution of elliptical galaxies induced by
          dynamical friction\\
          I. Capture of a system of satellites}

   \subtitle{}

   \author{G. Bertin,
          \inst{1,2}
          T. Liseikina,
          \inst{2,3}
          \and
          F. Pegoraro
          \inst{3}\fnmsep\thanks{Istituto Nazionale di Fisica della Materia,
          Sezione A}
          }

   \offprints{G. Bertin}

   \institute{Universit\`{a} degli Studi di Milano, Dipartimento di Fisica, via
             Celoria 16, I-20133 Milano, Italy\\
             \email{Giuseppe.Bertin@unimi.it}
         \and
             Scuola Normale Superiore,
              Piazza dei Cavalieri 7, I-56126 Pisa, Italy\\
              \email{t.liseikina@sns.it}
         \and
             Universit\`{a} degli Studi di Pisa, Dipartimento di Fisica, via Buonarroti 2,
             56100 Pisa, Italy\\
             \email{pegoraro@df.unipi.it}
             }

   \date{April 1, 2003, final version; originally submitted
 September 6, 2002}

   \abstract{The main goal of this paper is to set up a numerical laboratory
   for the study of the slow evolution of the density and of the pressure
   tensor profiles of
   an otherwise collisionless stellar system, as a result of the
   interactions with a minority component of heavier ``particles".
   The effects that we would like to study
   are those attributed to slow collisional relaxation and generically called
   ``dynamical friction"; in real
   cases, or in numerical simulations, the processes involved are
   complex, so that the relaxation associated with the granularity
   in phase space is generally mixed with and masked by evolution
   resulting from lack of equilibrium or from a variety of instabilities
   and collective processes. We start by revisiting the problem of
   the sinking of a satellite inside an initially isotropic,
   non-rotating, spherical galaxy, which we follow by means of
   N-body simulations using about one million particles.
   We then consider a quasi-spherical problem, in which the
   satellite is fragmented into a set of many smaller masses with
   a spherically symmetric initial density distribution.
   In a wide set of experiments, designed in order to bring out effects
   genuinely associated with dynamical friction, we are able to demonstrate
   the slow evolution of the density profile and the
   development of a tangentially biased pressure in the underlying
   stellar system, while we briefly
   address the issue of the circularization of orbits induced by
   dynamical friction on the population of fragments.
   The results of the simulations presented here
   and others planned for future investigations allow us to study the
   basic mechanisms of slow relaxation
   in stellar systems and thus may be of general interest for a variety
   of problems, especially in the cosmological context. Here our
   experiments are conceived with the specific goal to
   clarify some mechanisms that may play a role in the evolution
   of an elliptical galaxy as a result of the interaction between
   the stars and a significant population of globular clusters or
   of the merging of a large number of small satellites.
   }
\authorrunning{G. Bertin et al.}
\titlerunning{Evolution of galaxies induced by dynamical friction}
   \maketitle
%

\section{Introduction}

If we consider a globular cluster of mass $m_{gc} \sim 10^5
M_{\odot}$ orbiting inside a spherical galaxy, given the mass
ratio $10^5$ with respect to the mass of a typical star, we find
that it should suffer dynamical friction on a time-scale $T_{fr}
\sim 10^{9}$yr, through scattering of the stars. [The term
dynamical friction refers to the slow relaxation process
associated with discrete two-body encounters investigated in the
pioneering studies of Chandrasekhar (\cite{cha43}). In these
classical analyses, estimate is given of the cumulative effect of
discrete encounters on a {\it test particle} passing through a
homogeneous system of field particles. This effect, associated
with the granularity of the system in phase space, is ignored in
pure Vlasov descriptions of stellar systems. The term dynamical
friction is often extended to describe loosely the physical origin
of orbital decays of satellites of {\it finite mass} also in real
inhomogeneous systems and simulations, which fall obviously
outside the classical description of Chandrasekhar.] If the
cluster starts from a circular orbit located at $r_i$, in a time
$\sim T_{fr}$ it will reach a lower orbit close to $r_f$, while a
fraction of the energy initially associated with such a cluster
$E_i = m_{gc}\Phi(r_i)+ (1/2) m_{gc}r_i (d\Phi/dr)_{r_i}$ is
released into the distribution function of the hosting galaxy. If
we refer to the realistic possibility of the presence of several
thousands of such globular clusters we may thus argue that on a
fraction of the Hubble time the underlying stellar system may well
absorb a non-negligible fraction of its binding energy and thus
should slowly evolve (see also Lotz et al. \cite{lot01}).

Under the suggestive scenario just outlined, the issue we would
like to address in this paper is the following conceptually
challenging problem. In a two-component system, where the more
massive component (the ``galaxy") is basically collisionless and
the minority component (the globular cluster system or a large
number of mini-satellites caught in minor mergers) is slowly
dragged in toward the galaxy center by processes of dynamical
friction with the stars of the galaxy, how does the distribution
function of the galaxy evolve during the process? A resolution of
this challenging problem requires a clarification of the basic
mechanism of dynamical friction under inhomogeneous conditions and
a discussion of the collective effects involved in the reaction of
the underlying distribution function. Note that while the infall
of a single cluster (or satellite) is a problem with a preferred
direction (given by the angular momentum of the satellite), the
capture of a large number of randomly distributed clusters (or
mini-satellites) can preserve spherical symmetry, if present to
begin with.

It might be argued that the slow evolution in the quasi-isotropic
many-cluster case described above could be modeled in terms of an
{\it adiabatic evolution} of a distribution function within
spherical symmetry. For the different problem of the adiabatic
growth of a central massive black hole, such a procedure has
indeed been developed (e.g., see Cipollina \& Bertin \cite{cip94}
and references therein). Here it is not clear how a semi-analytic
procedure to calculate such slow evolution, taking advantage of
the conservation of the relevant actions, can be formulated. In
addition, given the relation between dynamical friction and {\it
resonant} interactions (e.g., see Weinberg \cite{wei89}), in spite
of its slowness the process is likely to be inherently
non-adiabatic. Therefore, it seems that the only viable way to an
answer is that of approaching the problem by means of suitable
N-body simulations.

In the long run, results from this project could also shed light
on another theoretically interesting scenario, which so far has
found major applications only in the context of the internal
dynamics of globular clusters. This is a picture, pioneered by
Bonnor (\cite{bon56}) in his analysis of gas spheres, by which
stellar systems may be subject to a process of gravitational
catastrophe (Lynden-Bell \& Wood \cite{lyn68}), when their central
concentration exceeds a certain threshold value. The catastrophe
consists in an instability process, where evolution leads to cores
that become more and more concentrated. The source of this
instability is thermodynamical, in the sense that it results from
a counter-intuitive property of self-gravitating systems, which
can be described in terms of negative specific heat. Recently,
this general scenario has been demonstrated to hold for an
astrophysically interesting family of partially relaxed stellar
systems (Bertin \& Trenti 2003).

The study of relaxation in N-body simulations and by means of
N-body simulations is a vast field of research (e.g., see Hohl
\cite{hoh73} and Huang et al. \cite{hua93}). The study of the
sinking of a single satellite inside a galaxy, as a result of
dynamical friction, was undertaken systematically about two
decades ago, initially through simplified simulations, within a
restricted three-body problem and with the center of mass of the
primary galaxy being kept fixed, and then by means of more
realistic, self-consistent simulations (Bontekoe \& van Albada
\cite{bon87}; Zaritsky \& White \cite{zar88}; Bontekoe
\cite{bon88}). The results showed that the general scaling
predicted by the classical formulae derived in the ``homogeneous
limit" (or ``local limit") (Chandrasekhar \cite{cha43}) was
basically applicable, although the sinking time scale turned out
to be significantly {\it longer} (by a factor of $2 - 3$ for the
regimes studied by the numerical experiments) if the
self-consistent response of the galaxy was properly incorporated.
A convincing interpretation of the sinking process can be made, at
least in the limit where the mass of the sinking satellite is
small, by investigating the linear response in the galaxy
distribution function through stellar dynamical techniques
(Fridman \& Polyachenko \cite{fri84}; Palmer \& Papaloizou
\cite{pal85}; Weinberg \cite{wei86}, \cite{wei89}; Bertin et al.
\cite{ber94}, hereafter BPRV94; Palmer \cite{pal94}). Such
response consists of an ``in-phase" contribution, which has little
relevance to the mechanism responsible for dynamical friction, and
an ``out-of-phase" contribution, associated with the ``resonant"
interaction, which produces a wake able to set a significant
torque on the satellite. These techniques demonstrate that the
collective wake in the galaxy, largely dominated by the dipolar
($l = 1$) response, can trail significantly away from the orbiting
object (when the satellite is outside the galaxy, the response is
concentrated around a location opposite to the object with respect
to the global center of mass), so that the resulting torque on the
satellite that is responsible for dynamical friction can be
significantly reduced with respect to other more naive
calculations. Discrepancies between earlier simulations and the
predictions of the stellar dynamic semi-analytic theory were
ascribed to non-linearity effects (in the sense that the
satellite-to-galaxy mass ratio $M_s/M_G \approx 0.1$ considered in
the simulations was judged to be excessive; see Hernquist \&
Weinberg \cite{her89}), but the general picture appears to have
been clarified. It is noted that the differences in behavior with
respect to the naive expectations of the traditional formulae for
dynamical friction become substantial when the size of the
satellite is finite (Weinberg (\cite{wei89}) refers to the length
ratio $R_s/R_G$; the multipoles associated with the perturbation
with $l$ above a certain threshold are bound to be unimportant),
while the differences should become negligible for very compact
satellites (in which case all multipoles contribute, with a
divergence that is removed by the Coulomb logarithm).


Several aspects of the general problem of dynamical friction have
been revisited recently. The issue of the contrast between
``local" and ``non-local" effects has been re-discussed (Prugniel
\& Combes \cite{pru92}; Maoz \cite{mao93};  Cora, Muzzio \& Vergne
\cite{cor97}), with the confirmation that the description of the
friction process in terms of the ``homogeneous limit"
(Chandrasekhar \cite{cha43}) should be adequate in the limit of
low-mass, relatively compact satellites. A very recent
investigation (Cora, Vergne \& Muzzio \cite{cor01}) also suggests
that the process depends very little on the regularity or the
stochasticity of the stellar orbits in the galaxy, in contrast
with some earlier conjectures (Pfenniger \cite{pfe86}). Other
interesting problems are the problem of circularization (of the
satellite orbit), the relation of the friction effect to the past
history of the dynamical event under consideration, and the issue
of a direct calculation of the relevant dissipative force from the
response of the stellar system to the perturbing satellite (e.g.,
see S\'{e}guin \& Dupraz \cite{seg94,seg96}; see also Colpi
\cite{col98}; Colpi \& Pallavicini \cite{col98a}; Colpi, Mayer \&
Governato \cite{col99}; Nelson \& Tremaine \cite{nel99}).

Note that so far investigations of processes of dynamical friction
have mostly focused on the issue of the fate of the sinking
objects (and mostly for the case of one object; but see Tremaine,
Ostriker \& Spitzer \cite{tre75}), while the problem of the
reaction of the distribution function of the underlying stellar
system (especially in the quasi-spherical limit of a system of
many heavy objects outlined above) remains practically unexplored.

Our paper is organized as follows. In Sect.~\ref{sect:theory}
we summarize the
theoretical framework that allows us to understand the mechanisms
of dynamical friction within the context of a single sinking
satellite. In Sect.~\ref{sect:model} we describe the model adopted. For the
present exploratory investigation, the choice of the model (both
for the galaxy and for the satellite or the fragments) is mostly
dictated by convenience (for the study of dynamical mechanisms and
for an easier comparison with previous analyses) rather than by
realistic requirements. In Sect.~\ref{sect:tests} we describe the diagnostic
tools that will be used in the simulations and some preliminary
tests. In Sect.~\ref{sect:results} we proceed to illustrate the results of the
simulations for the case of one sinking satellite and for the
quasi-spherical case for which the sinking of many smaller
mini-satellites is considered. In Sect.~\ref{sect:conclusion} we briefly draw our
conclusions and set out the plans for future investigations.
\label{sect:intro}
\section{Theoretical framework for the problem of a single sinking satellite}
\label{sect:theory}
Here we briefly recall the main issues involved in the problem of
a single sinking satellite of ``diameter" $2 R_s$, initially
located on a circular orbit at distance $r_s \approx R_G$ from the
center of mass of the entire system. Here $R_G$ denotes the
``radius" of the galaxy. Let $\Omega_s > 0$ be the angular
velocity of the satellite along its circular orbit. The case of a
very compact satellite is likely to be within the reach of the
classical formulae for dynamical friction (Chandrasekhar
\cite{cha43}). In the case where the satellite is sufficiently
extended, for the description of its density distribution and the
associated gravitational potential, only the first harmonics in a
multipole expansion are important, up to $l_{max} \sim \pi r_s/(2
R_s)$ (see Weinberg \cite{wei89}).

A ``friction" torque acting on the satellite is expected to result
from the action of suitable resonances with star orbits. To
understand this, we may refer to the action of the satellite as a
perturbation over a basic state that is non-rotating, spherically
symmetric, and characterized by isotropic distribution function
$f_0 (E)$, where $E = v^2/2 + \Phi_0 (r)$ is the star energy, per
unit mass, in the absence of the perturbation, and we may then
follow the analysis developed earlier
(BPRV94).

Analytical results can be derived in the limit when the mass of
the satellite is vanishingly small (with respect to the mass of
the galaxy, so that the resulting perturbation in the galaxy
distribution function can be treated by a linear theory; see
subsection \ref{sect:shield}). In practice, for real systems or for numerical
simulations, some of the evolution effects are going to be
associated with the {\it finite mass} of the satellite.
Once we move to consider the possibility of a quasi-equilibrium
configuration for a composite system (galaxy plus satellite or a
population of mini-satellites), we then have to worry about
genuine instabilities that may occur, independent of the
granularity present in the system. All of this will contribute to
evolution although, in principle, it is not related to the
original {\it gedanken experiment} at the basis of the classical
definition of dynamical friction.

Another subtle aspect of the problem is related to geometry.
Broadly speaking, we may think that the difference between the
case of straight orbits (considered in the classical description
of dynamical friction) and that of orbits in the inhomogeneous
potential of the galaxy is taken into account in the following
discussion of the resonant response. Yet, the cumulative effects
on a {\it single star}, because of multiple encounters with the
satellite (for example, for stars on quasi-circular orbits), grow
in time differently from the classical picture, even before
dealing with the collective behavior of the entire set of stars
that characterizes the resonant response.

\subsection{Single-star Resonances}
\label{sect:single_star}
The frequency $\omega$ associated with the $m$-th component of a
multipolar term of order $l$ in the expansion of the potential
$\Phi_s$ of the satellite is readily related to the angular
frequency of the satellite $\omega = - m \Omega_s$. Therefore, if
we refer to Eq.~(80) of BPRV94, we find that the appropriate
resonance condition is $ - m \Omega_s - p \Omega_r(E,J) + s
\Omega_{\theta}(E,J) = 0$; here $s$, $m$, $p$ are integers ($- l
\le s,m \le +l$; $- \infty < p < \infty$) and $\Omega_r(E,J) > 0$,
$\Omega_{\theta}(E,J) > 0$ are the two frequencies (dependent on
the star specific energy $E$ and angular momentum $J$) that
characterize the star orbit and depend on the properties of the
basic state. As discussed in BPRV94, only the components with $s =
l, l-2, .. -l$ contribute to the density perturbation.

For an extended satellite, the $p = 0$ component, which
corresponds to an average along the star radial orbit excursion,
is expected to dominate so that the resonance condition should
reduce to $\Omega_{\theta}(E,J) = m \Omega_s/s$.
Note that in the course of time, because of dynamical friction,
the angular frequency of the satellite is going to change.
Clearly, the resonance conditions that we have just stated are
applicable provided that $\dot{\Omega}_s/\Omega_s \ll
\Omega_{\theta}(E,J)$, which we can rewrite as $\dot{\Omega}_s \ll
\Omega_s^2/l$.


\subsection{``Shielding" and global resonances}
\label{sect:shield}
If we consider the linearized Vlasov-Poisson equation for our
problem, we see that Eq.~(54) of BPRV94 is modified into $f_1 =
\mathcal{L}(\Phi_s + \Phi_1)$, where $\mathcal{L}$ is a linear
operator that basically describes one integration along the
unperturbed characteristics. The quantity $f_1$ can be called the
linear response of the distribution function. This response has a
``driven" part, associated with $\Phi_s$, and a
``self-gravitating" part, related to the potential $\Phi_1$, that
must be consistent with the response itself, following the Poisson
equation $\nabla^2 \Phi_1 = 4 \pi G \int f_1 d^3 v$.

\noindent In plasma physics, the latter contribution is often
called the ``shielding" associated with the collective behavior of
the plasma. Such shielding is often ignored, although it should be
emphasized that this step is {\it improper}, because the omitted
term is a {\it linear} contribution. A relatively simple analysis
can then be made, by neglecting the self-gravity of the response,
showing the connection between {\it single-star resonances} and
dynamical friction. Note that this is the only ground where a
comparison with Chandrasekhar's (\cite{cha43}) work can be made.

The general case, which includes the possibility of resonance with
{\it discrete global modes} of the system, is currently beyond the
reach of analytical investigations (see also Vesperini \& Weinberg
2000).

\section{Model}
\label{sect:model}
As a model of the primary stellar system (the ``galaxy"),
following Bontekoe \& van Albada (\cite{bon87}) we consider an
isotropic polytropic model of index $n = 3$. The distribution
function is thus of the form $f_G(\vec{r},\vec{v}) =
A(E_0-E)^{n-3/2}$ for $E<E_0$, with $f_G(\vec{r},\vec{v}) = 0$ for
$E>E_0$, where $E = \Phi_G(r) +  v^2/2$  {\rm is the single-star
energy}, $E_0 = \Phi_G(R)  = - GM/R $ is the gravitational
potential at the boundary $r=R,$ and $M$ is the total mass of the
system. Polytropes can be described in terms of the Lane-Emden
function $\theta(\xi),$ which satisfies the differential equation:

\begin{equation}
\label{eq:emd} \frac{1}{\xi^2} \frac{d}{d \xi} (\xi^2 \frac{d
\theta}{d \xi}) = - \theta^n,
\end{equation}

\noindent where $n$ is the polytropic index, with the boundary
conditions $\theta = 1$, $d \theta / d \xi = 0$ at $\xi = 0$. From
the function $\theta(\xi)$ one can find the potential $\Phi_G(r)$
and the associated density $\rho_G(r)$.


The secondary object (the ``satellite") is described by a rigid
Plummer density distribution (corresponding to a polytrope of
index $n = 5$), characterized by $\theta = (1 + 0.5
\xi^2)^{-1/2}$. The potential and the mass distribution of the
satellite are then defined by two scales, $R_s$ and mass $M_s$, so
that $\Phi_s (r) = - G M_s (R_s^2 + r ^2 )^{-1/2}$, with
cumulative mass $M_s(r) = M_s r^3(R_s^2 + r^2 ) ^{-3/2}$. This
distribution extends to infinity, but 90\% of the mass is
contained within $3.7 R_s.$

As a result of the interaction with the satellite, the
distribution function of the galaxy is a time-evolving $f$ that
will be associated with a mean field $- \nabla \Phi$, to be
derived from the Poisson equation $\Delta\Phi(\vec{r},t) =  4 \pi
G \int f d^3 v = 4 \pi G \rho(\vec{r},t)$. We solve the Poisson
equation on a spherical polar mesh of $24\times 12\times 24$ cells
in the $r,\theta$, and $\varphi$ directions. The radial mesh is
logarithmic. The angular mesh is fixed, with 12 equal divisions in
$\cos\theta$ and 24 in $\varphi.$ On this mesh the density, the
potential, and the force field are expanded in associated Legendre
functions $P_l^m(\cos\theta)$ up to $P_6^6.$ At each time-step the
center of the spherical mesh used to calculate the potential and
the force field in the galaxy is repositioned so as to coincide
with the center of mass of the galaxy. The equations of the motion
(for the particles representative of the galaxy and, separately,
for the satellite) are integrated in Cartesian coordinates with
the so called leapfrog scheme. This is basically the N-body code
described by van Albada \& van Gorkom (\cite{van77}) and then
improved and used by van Albada (\cite{van82}) and by Bontekoe \&
van Albada (\cite{bon87}). Therefore, we do not provide here a
full description of the method employed, since it is available in
the papers quoted.

\subsection{Sinking of a single satellite}
\label{sect:single}
A first set of numerical experiments addresses the sinking of a
single satellite, initially located on a circular orbit at $\theta
= \pi/2$.

The galaxy is considered as a collisionless stellar system, so
that its representative particles $(i = 1,....., N)$ interact with
each other through the mean gravitational potential $\Phi$ and
directly with the satellite:

\begin{equation}
\label{eq:motG} \frac{d^2 \vec{r}_i}{d t^2} = -
\nabla\Phi(\vec{r}_i,t) - \frac{G M_s
(\vec{r}_i-\vec{r}_s)}{[R_s^2 + (\vec{r}_i-\vec{r}_s)^2]^{3/2}}.
\end{equation}

\noindent Note that $\Phi$ is updated separately by the
Poisson-solver scheme, following the original method described by
van Albada \& van Gorkom (\cite{van77}) (this step effectively
closes the system of equations), which guarantees a pure
time-reversible evolution.

In turn, the satellite is taken to interact with all the
representative particles of the galaxy directly via two-body
forces:

\begin{equation}
\label{eq:mots} \frac{d^2 \vec{r}_s}{dt^2} = \sum_{i=1}^{i=N}
\frac{G m (\vec{r}_i - \vec{r}_s)} {\left[R_s^2 + (\vec{r}_i -
\vec{r}_s)^2\right]^{3/2}},
\end{equation}

\noindent In the equations above, $\vec{r}_i, m$ are the position
vectors and the mass of the representative particles of the
galaxy, while $\vec{r}_s$ is the position vector of the satellite.
All position vectors are measured with respect to an inertial
frame of reference with its origin at the center of mass of the
combined (galaxy + satellite) system.

\subsection{Sinking of many fragments}
\label{sect:many}
Another set of experiments studies an initial configuration where
the satellite is fragmented into several ($N_f$) pieces
(``mini-satellites" or ``fragments"). For simplicity, we take the
fragments to be characterized by the same Plummer density
distribution with mass $M_f$ (such that $M_s = N_f M_f$) and the
same lengthscale $R_f = R_s$ (a smaller lengthscale $R_f$ would be
more realistic for the astronomical system we have in mind, but
would be less easy to simulate).

The mutual interactions between mini-satellites can be kept ``on''
by considering

\begin{equation}
\label{eq:motGN} \frac{d^2 \vec{r}_i}{dt^2} = -
\nabla\Phi(\vec{r}_i,t) - \sum_{j=1}^{j=N_f} \frac{G {M_f}
(\vec{r}_i - \vec{r}_{fj})} {[R_f^2 + (\vec{r}_i -
\vec{r}_{fj})^2]^{3/2}},
\end{equation}

\noindent for the representative particles of the galaxy (labeled
by $i = 1, ...., N$), and

\begin{equation}
\label{eq:motsN} \frac{d^2 \vec{r}_{fj}}{dt^2} = \vec{a}_j^{(1)} +
\vec{a}_j^{(2)}
\end{equation}

\noindent for the mini-satellites (labeled by $j = 1,..., N_f$),
where

\begin{equation}
\label{eq:motsN1} \vec{a}_j^{(1)} = \sum_{i=1}^{i=N} \frac{G m
(\vec{r}_i - \vec{r}_{fj})} {\left[R_f^2 + (\vec{r}_i -
\vec{r}_{fj})^2\right]^{3/2}}
\end{equation}

\noindent is the acceleration produced by the galaxy and
$\vec{a}_j^{(2)}$ denotes the gravitational acceleration exerted
on the $j$-th mini-satellite by the other fragments. The
gravitational interaction between two extended bodies is
complicated to describe. For the purposes of the present paper,
where we are not interested in the phenomena associated with the
details of such interaction, we simplify the simulations by
referring to the following prescription:

\begin{equation}
\label{eq:motsN2} \vec{a}_j^{(2)} = \sum_{k =1,k \ne
j}^{k=N_f}\frac{G M_f (\vec{r}_{fk} - \vec{r}_{fj})} {\left[R_f^2
+ (\vec{r}_{fk} - \vec{r}_{fj})^2\right]^{3/2}}.
\end{equation}

\noindent All position vectors are measured with respect to an
inertial frame with its origin at the center of mass of the entire
system.

The mutual interaction between mini-satellites can be ignored and
turned ``off" simply by taking $\vec{a}_j^{(2)} = 0$. This
procedure may be the more realistic procedure if we wish to
simulate a system of globular clusters inside a galaxy. In fact,
for such system, the few-body-problem effects associated with such
interaction are expected to play a secondary role.

As to the initial distribution of the fragments we have considered
three cases, to be described below. The first is the most natural
generalization of the study of the sinking of a single satellite.
A posteriori, we have found that in such a case the effects
related to the lack of equilibrium in the initial conditions are
too strong and confuse the picture of evolution. (Similar
problems, related to the lack of equilibrium, were present also in
the case of a single satellite, but those are not emphasized in
this paper because that case has been well studied in the past and
we prefer to focus here on the new quasi-spherical problem in the
presence of many fragments). Therefore we have devised quieter
starts so as to better disentangle the effects associated with the
granularity of the system from those associated with the lack of
equilibrium.

Given the relatively small number of fragments involved, we
anticipate that different realizations of the physical cases
described below may be associated with different evolutions.

\subsubsection{I: Fragments quasi-isotropically distributed on
circular orbits on a thin shell}
\label{sect:case1}
We first consider the case where all the fragments are distributed
initially with random positions on a thin shell, located at radius
$r_{shell}(0)$. The fragments are given initial velocities
appropriate for circular orbits, with random orientations, based
on the gravitational field generated by the galaxy alone.

The purpose of simulations of this type is to derive information
relevant for a more realistic situation where the fragments have a
full three-dimensional space distribution, so that the interaction
between the galaxy and the minority population is differential
with radius. Such ideally desired simulations are not feasible
with current codes. To move one step further in the desired
direction we have then considered the case of two shells of
fragments initially located at different radii.

Unfortunately, due to the finite mass of the set of fragments,
these simulations are characterized by violent epicyclic
oscillations that shake the system on the fast dynamical time
scale and tend to persist long after the initially expected
transient.

\subsubsection{II: Fragments quasi-isotropically distributed on
circular orbits on a thick shell with improved initial conditions}
\label{sect:case2}
In order to set up a smoother quasi-equilibrium initial condition,
we have decided to proceed as follows. We refer to a spherically
symmetric density distribution for a shell of the form
$\rho_{shell} = \rho_0 \exp[-(r-r_{shell}(0))^2/a^2]$ for $
|r-r_{shell}(0)| \le 2 a$ (and vanishing for $|r-r_{shell}(0)|> 2
a$). The normalization constant $\rho_0$ is taken in such a way
that the total mass associated with the shell is $M_s$. In view of
the choice of model for the fragments, we have decided to take $2
a = R_s$.

Such density distribution generates a potential $\Phi_{shell}(r)$,
which can be calculated numerically. We now consider a modified
distribution function for the galaxy $f_G(\vec{r},\vec{v})$ taken
to be of the polytropic form as in Sect.~\ref{sect:model}, but with $E =
\Phi_G(r) + \Phi_{shell}(r) +v^2/2$. We then have to integrate a
Lane-Emden equation similar to Eq.~(\ref{eq:emd}) but modified by
the presence of the external potential $\Phi_{shell}(r)$. This
yields a potential $\Phi_G(r)$ different from that considered in
Sect.~\ref{sect:model}, to be inserted into the definition of
$f_G(\vec{r},\vec{v})$.

Finally, we consider the clumpy realization of the shell density
distribution by reducing it to a set of $N_f$ fragments of mass
$M_f$. The $j-$th fragment is initially distributed in $r$ at a
location $r_{fj}$ (with random angular coordinates) so that the
set of fragments reproduces the assumed density distribution of
the shell; its initial velocity is that appropriate for the
circular velocity of a test particle in the combined potential
$\Phi_G(r) + \Phi_{shell}(r)$.

This procedure ensures a quieter start while leaving the basic
picture of the single shell case unchanged. It can be generalized
to the case of two shells of fragments initially located at
different radii.

\subsubsection{III: Fragments constructed by clumping part of the
galaxy distribution function}
\label{sect:case3}
The cases considered above are aimed at describing the interaction
between the galaxy and a minority component with phase-space
properties distinct from those of the galaxy. As we have seen
while addressing the need for a quieter start, such distinction is
likely to be accompanied by effects that are due to the lack of
equilibrium in the initial conditions, difficult to separate from
the secular effects that are traditionally called dynamical
friction. In this assessment, we can go even one step further and
we may argue that some of these initial conditions (with the
different phase-space properties considered) may actually be
characterized by dynamical instabilities. If this were the case,
it would be very hard or even impossible to distill the true
effects of dynamical friction out of the simulations. In order to
address this point, we will run parallel simulations of case II
with a smooth, Vlasov realization of the shell density, which
should allow us to single out the effects of possible Vlasov
instabilities of such system (see Subsect.~\ref{sect:collisionless}).

In view of this discussion, we have decided to address a third set
of simulations that is less relevant for the astrophysical case
that has motivated this paper but is definitely more interesting
for the study of the problem of dynamical friction.

Here we consider the initial equilibrium state to be basically
that described by the galaxy model {\it alone, without satellites
or shells or fragments}, as defined in the first paragraph of
Sect.~\ref{sect:model}.
The mini-satellites here are then introduced by clumping
together part of the distribution function $f_G$ in such a way
that it corresponds to a diffuse ``shell" in space. Formally, we
separate out a piece of the distribution function $f_{shell}$ (for
simplicity, we keep the notation ``shell" even if we are
considering a situation different from that of case I and case
II). Then the system made of $f_G - f_{shell}$ is simulated by the
Vlasov code, while the contribution $f_{shell}$ is reduced to
clumps that are brought back to the form of the fragments and
treated by means of direct interaction with the galaxy
representative particles (as described in subsection~\ref{sect:many}).
Note
that in this case the ``clumps" or ``fragments" are distributed in
phase space exactly as the particles of the galaxy (thus
minimizing the lack of equilibrium in the initial conditions and
the risk of introducing undesired sources of instability); in
particular, the clumps are initially characterized by an {\it
isotropic} distribution in velocity space.

We should emphasize that this third class of simulations, while
addressing the settling of heavy masses dragged inwards by
dynamical friction, is conceptually distinct from the study of the
mass stratification process in self-gravitating systems, which may
be performed by means of direct or Fokker-Planck codes (e.g., see
Spitzer \cite{spi87}).

\subsubsection{Purely collisionless reference models}
\label{sect:collisionless}
For the three cases just outlined, the simulations carried out
following the description given in subsection~\ref{sect:many} will be
accompanied by reference simulations of similar but purely
collisionless models. These can be seen as cases for which the
number of fragments formally diverges, while the total mass
contained in the fragments remains constant. In the simulations
the case of very large number of fragments cannot be treated in
terms of direct interaction; instead, for these cases, the
fragments are treated as simulation particles of the collisionless
component. It is clear that for case III, in such reference
collisionless models, the identification of $f_{shell}$ is
meaningless. In practice, to monitor the properties of our
integration scheme, we have labeled the representative particles
associated with $f_{shell}$ by a different ``color".

The purpose of these parallel reference runs is to check, as much
as possible, that we are not confusing evolution effects
associated with the initial conditions with the effects that are
genuinely associated with the granularity of the system.

\subsubsection{Initial set up}
\label{sect:setup}
In order to derive the initial distribution of the particles in
the simulation we start from the mass distribution $M(r)$ which is
known on a discrete set of points labelled by the index $j$, with
$M_j = M(r^j)$.

First we determine the distance of a particle from the galaxy
center. If all the particles of the galaxy have the same mass $m =
M/N$, then for the $i^{th}$ particle we define $xx = m \cdot (i -
1/2),$ and then find the cell $j$ such that $M_{j}>xx>M_{j-1}.$
The distance from the center $r_i$ of the $i^{th}$ particle is
then found by an interpolation between $r^{j-1}, r^j$, and
$r^{j+1}$. Next we choose two uniformly distributed  random
numbers: $q_{\varphi}\in[0,1]$ for the $\varphi$ coordinate of a
particle, such that $\varphi=2\pi q_{\varphi}$, and
$q_{\theta}\in[-1,1]$ for the $\theta$ coordinate, such that
$\cos\theta = q_{\theta}.$ The Cartesian coordinates are then
found from the spherical coordinates by standard formulae. The
choice of the initial velocities is performed in a similar way,
using a rejection method (e.g., see Press et. al. (\cite{pre92}))
to obtain the desired distribution function. For the potential
$\Phi_G(r)$ we proceed as for the mass distribution; the value of
the potential at the radius of the particle location is found by
interpolation between the values at $r^{j-1}, r^j$, and $r^{j+1}.$

In Case III, of a ``subtracted" shell, the particles of the galaxy
are divided in two ``species": light particles, with mass $m= M/N
=(M-M_s)/N_1$, and heavy particles with mass $M_f = M_s/N_f$
respectively, where $M_s$ is the mass of the shell, $N_f$ the
number of heavy particles, and $N_1$ the number of light
particles. Initially, the heavy particles are all located inside a
thin shell. Inside this shell there are no light particles. The
$i^{th}$ light particle, for $i\in[1,i1=N\cdot M(r_{shell})/M],$
where $r_{shell}$ is the position of the ``shell", is located  as
in the non-subtracted case at the distance $r_i$ obtained by
interpolation between the $r^{j-1}, r^j$, and $r^{j+1},$ where $j$
is such that the condition $M_{j}>m\cdot (i-1/2)>M_{j-1}$ is
satisfied. Then, for the  $k^{th}$ heavy particle, with
$k\in[1,N_f]$, we adopt $xx = m \cdot (i1 - 1/2) +  M_f \cdot
(k-1/2) $ and then find the cell $j$ where $M_{j}>xx>M_{j-1}.$ The
distance from the center $r_k$ of the $k^{th}$ heavy particle is
then found by the interpolation rule mentioned above. After that
the remaining light particles are distributed according the
previous rule. Finally, in the simulations, the light particles
interact with one another through their mean field, while the
heavy particles interact directly with the light particles and
among themselves, as described at the beginning of Sect.~\ref{sect:many}.

\subsubsection{Other possible initial conditions}
\label{sect:other}
Another possible initial condition is to turn on slowly  the
gravity associated with the massive objects or massive shells. In
a  collisionless simulation, it would lead to the ``empirical
construction" of equilibrium systems in which dynamical friction
is absent and would thus be, for our purposes, essentially
equivalent to  the procedure described in Case II.  Such
construction of equilibrium models can also be approached
analytically (see Cipollina \& Bertin \cite{cip94} and references
therein).

The case where the satellite or the fragments are treated as
discrete objects would provide information on the friction force,
different from the information provided in this paper, but such
approach would have no special merits, in terms of physical
justification and physical interpretation, with respect to the one
adopted here.

\subsection{The choice of the code}
\label{sect:choice}

The choice of code that we have made (basically, the code used by
Bontekoe \& van Albada (\cite{bon87})) appears to be appropriate
for the study of the effects of dynamical friction, which are
dominated by low-$l$ wakes (see Sect.~2). Still, one may ask the
reason why we have resorted to that code instead of codes that are
commonly utilized for a variety of studies in galactic dynamics or
in the cosmological context. In particular, one popular class of
codes, the {\it tree-codes} (e.g., see Barnes \& Hut
(\cite{bar86})), has found modern realizations (e.g., GADGET; see
Springel, Yoshida, \& White \cite{spr01}) that are particularly
appealing since they are flexible, widely used, and have gone
through many tests. Tree-codes treat near particles in terms of
direct interactions via a {\it softened} potential, because the
use of the true Newtonian potential would introduce a generally
undesired collisionality among the simulation particles. Tree
codes are often used to study the evolution of collisionless
systems; for this purpose, the softening length has to be chosen
in an optimal way, depending on the physical characteristics of
the system that is being investigated.

The use of a tree code in our project would then pose severe
interpretation problems, because the non-physical effects
associated with the introduction of the softening parameter would
show up precisely in relation to the issues that we are trying to
investigate.



\section{Diagnostics and test cases}
\label{sect:tests}
The results of our simulations can be studied in terms of the
energy and angular momentum budget as a function of time and by
means of figures that illustrate the spatial structure of the
density and of the pressure distribution within the galaxy in the
course of time. Another set of figures will provide plots of the
azimuthally averaged density profile for the galaxy and of the
location $r_s$ of the satellite as a function of time; for the
case of many mini-satellites, we will show a plot of the radius
$r_{x}$ of the sphere that contains different fractions ($x$
percent) of the total mass in the form of mini-satellites. We
first proceed by specifying units and definitions for the relevant
quantities.

\subsection{Units}
\label{sect:units}
Unless specified otherwise, the units adopted for mass, length,
and time are the mass of the galaxy ($M$), the radius of the
galaxy ($R$), and the natural crossing time ($t_{cr} = G M^{5/2}(2
K_{gal})^{-3/2}$ evaluated at the beginning of the simulation;
here $K_{gal}$ is the total kinetic energy associated with the
stars in the galaxy). For the galaxy model considered in this
paper, the revolution period relative to a circular orbit at the
periphery (at $R$) is $\approx 11~t_{cr}$. To return to physical
units, we may refer to the case where $M = 2 \cdot
10^{11}~M_{\odot}$, $R = 10$ kpc, so that $t_{cr} = 0.18 \cdot
10^8$ yrs and the revolution period at the periphery is $\approx 2
\cdot 10^8$ yrs.

\subsection{Conservation laws}
\label{sect:conservation}
In the course of the simulation the total energy and the total
angular momentum should remain constant at their initial values
$E_{tot}$ and $\vec{J}_{tot}$.

In the case of a single satellite we write the energy conservation
as

\begin{equation}
\label{eq:mots0} E_{tot}= E_{gal} + K_{sat} + E_{int}=
\end{equation} $$\sum_{i=1}^{i=N} \frac{m}{2} \left[
{\vec{v}_i^2} +\Phi(\vec{r}_i) \right] + \frac{M_s}{2}\vec{v}_s^2
- \left[\sum_{i=1}^{i=N} \frac{G M_s m}{(R_s^2
+(\vec{r}_i-\vec{r}_s)^2)^{1/2}}\right]~,$$

\noindent where $E_{gal} = K_{gal} + W_{gal}$ is the total
(kinetic plus gravitational) energy of the galaxy in the absence
of the satellite. From the point of view of the satellite, it is
natural to introduce the energy $E_{sat} = K_{sat} + E_{int}$, so
that we may split the total energy into $E_{tot}= E_{gal} +
E_{sat}$. The angular momentum conservation is given simply by
$\vec{J}_{tot}=  \vec{J}_{gal} + \vec{J}_{sat} =\sum_{i=1}^{i=N}
m\,\vec{r}_i \times \vec{v}_i + M_s \, \vec{r}_s\times
\vec{v}_s~$.

In the case of several mini-satellites, following the model
outlined in Sect.~\ref{sect:model}, the energy takes the form

\begin{equation}
\label{eq:mots000} E_{tot}= E_{gal} + (K_{fra} + W_{fra}) +
E_{int}= \sum_{i=1}^{i=N} \frac{m}{2} \left[ v_i^2
+\Phi(\vec{r}_i) \right] + \end{equation} $$ \sum_{j=1}^{j=N_f}
\frac{M_f}{2}\left[ v_{fj}^2 - \sum_{k\not= j}^{k=N_f}\frac{G
M_f}{(R_f^2 +(\vec{r}_{fj}-\vec{r}_{fk})^2)^{1/2}}\right] $$ $$-
\sum_{i=1}^{i=N} \sum_{j=1}^{j=N_f} \frac{G m M_f}{(R_f^2
+(\vec{r}_i-\vec{r}_{fj})^2)^{1/2}}~,$$

\noindent while the angular momentum can be written as
$\vec{J}_{tot}= \vec{J}_{gal} + \vec{J}_{fra} = \sum_{i=1}^{i=N}
m\, \vec{r}_i \times \vec{v}_i + \sum_{j=1}^{j=N_f}M_f \,
\vec{r}_{fj} \times \vec{v}_{fj}$. Note that the energy associated
with the system of mini-satellites now includes a term $W_{fra}$
that describes the gravitational energy internal to the system of
fragments. By analogy with the case of the single satellite, we
may wish to refer to the energy $E_{fra} = (K_{fra} + W_{fra}) +
E_{int}$ as the total energy associated with the fragments, so
that $E_{tot} = E_{gal} + E_{fra}$.

\subsection{Density and pressure distributions}
\label{sect:pressure}
In order to describe the galaxy density perturbation, we refer to
the density response defined as $\rho_1 (t,\vec{r}) =
\rho(t,\vec{r}-\vec{r}_0(t))-\rho_0(\vec{r})$; here
$\rho_0(\vec{r})$ is the adopted density distribution of the
galaxy in the absence of satellites, $\vec{r_0}(t)$ is the
position of the center of mass of the galaxy in an inertial frame
of reference. To diagnose the structure of the response it will be
useful to consider the characteristics of the first multipoles.

In the simulations the density on the spherical mesh is expanded
in associated Legendre functions $P_l^m(\cos\theta)$ up to $l=6$

\begin{equation}
\label{eq:mots1} \rho(r,\theta,\varphi) = \sum_{l=0}^{l=6}
{\bigg[}A_{l,0} (r)\, Y_l^0(\cos\theta)+\sum_{m=1}^{m=l} \sqrt{2}
A_{l,m}(r)\, {\rm Re} (Y_l^m(\cos\theta,\varphi))
\end{equation}
$$+\sum_{m=1}^{m=l} \sqrt{2} B_{l,m}(r)\, {\rm Im}
(Y_l^m(\cos\theta,\varphi)){\bigg]}~,$$

\noindent where the coefficients of the expansion into real
spherical harmonics are given by suitable integrations over the
relevant solid angle

\begin{equation}
\label{eq:mots2} A_{l,m}(r) = \sqrt{2} \int \rho(r,\theta,\varphi)
{\rm Re}(Y_l^m(\cos\theta,\varphi)) d\Omega~,
\end{equation}

\begin{equation}
\label{eq:mots3} A_{l,0}(r) = \int\rho(r,\theta,\varphi)
Y_l^0(\cos\theta) d\Omega~,
\end{equation}

\begin{equation}
\label{eq:mots4} B_{l,m}(r) = \sqrt{2} \int \rho(r,\theta,\varphi)
{\rm Im}(Y_l^m(\cos\theta,\varphi)) d\Omega~.
\end{equation}

\noindent In particular, the dipole ($l=1$) component of the
density response $\rho_1$ is given by

\begin{equation}
\label{eq:mots5} \rho_{l=1}(r,\theta,\varphi)=A_{1,0}
\sqrt{3/4\pi}\cos\theta - A_{1,1}
\sqrt{3/4\pi}\sin\theta\cos\varphi - B_{1,1}
\sqrt{3/4\pi}\sin\theta\sin\varphi~.
\end{equation}

The galaxy pressure tensor is defined in terms of the star
distribution function in the standard way.
In the case of a single satellite sinking into a galaxy along the
equatorial plane we have $\langle v_r \rangle \simeq 0$ and
$\langle v_{\theta} \rangle \simeq 0.$

\subsection{Tests}
\label{sect:test1}
We have performed many preliminary experiments aimed at testing
the numerical code.


The numerical implementation of the polytropic equilibrium
configuration of Sect.~\ref{sect:model} with $N=5 \cdot 10^3 \div 5 \cdot
10^5$ particles has been tested by generating an unperturbed
galaxy that has remained quiescent for about 50 crossing times.
The radii of the spheres containing $0.5 \%, 1 \%, 1.5 \%, ...., 5
\%$ of the total mass have been checked to remain constant in time
to within $1 \%$.

Then, we have tested that a satellite with $R_s = 0.1~R$ and very
small mass, $M_s = 10^{-8}~M$, behaves as a light ``test particle"
without disturbing the main galaxy. This satellite, placed
initially at the periphery of the galaxy $r_s = R$ on an orbit
with $v_{s\varphi} = 0.536~R/t_{cr}$ and $v_{sr} = 0.954 \cdot
10^{-3}~R/t_{cr}$, remains very close to its initial circular
path, never exceeding an eccentricity of $e=0.05$. At the end of
the run, at time $t \sim 50~t_{cr}$ the orbital energy of this
light satellite is conserved to within $1\%$.

\subsubsection{Energy and momentum conservation}
\label{sect:energy_conservation}
In addition to the specific tests mentioned above, we have checked
that globally energy and angular momentum are conserved, at $t
\sim 50~t_{cr}$, to better than $1\%$.

\subsubsection{Comparison with earlier work by Bontekoe and van Albada (1987)}
\label{sect:VanAlbada}
The main results of the simulations of the sinking of a single
satellite by Bontekoe \& van Albada (1987) have been checked in
detail, with the use of simulations with $N$ up to $5 \cdot 10^5$
(see Sect.~\ref{sect:single_result} below).
This adds much confidence, in relation both
to the overall structure of the code used in this paper and to the
modeling of dynamical friction used in these investigations (since
results have been checked to be largely independent of $N$ over a
range significantly larger than that available earlier).

\section{Results of the simulations}
\label{sect:results} We have performed a number of runs, mostly
based on $N = 5 \cdot 10^5$. Runs labeled by $A$ refer to
experiments with a single sinking satellite, initially located on
a circular orbit at $r_s = R$; various combinations of satellite
mass $M_s/M$ and satellite size $R_s/R$ have been considered. For
these runs (some are listed in Table~\ref{tab:tab2a}), we have
recorded the time $t_{fall}/t_{cr}$ at which the satellite reaches
the central regions; for some experiments (in particular, for
those labeled $A_8$ and $A_9$) the satellite is unable to reach
the center and we have thus studied the radius of minimum radial
approach $r_{min}$.

Runs labeled by $B$ are listed in Table~\ref{tab:tab2b}
and refer to the case of
the slow evolution of the system in the presence of many
fragments, initially located in a single shell, following the
procedure of case I described in Sect.~\ref{sect:case1}.
Runs of type $BS$ are
based on the smoother initialization of case II outlined in
Sect.~\ref{sect:case2}. Runs of type $BT$ refer to
the procedure of case III of
Sect.~\ref{sect:case3}.
Here the fourth column records the initial radius of the
shell and the fifth column lists the time at the end of the
experiment; all runs are made with the mutual interaction, among
fragments, on, but we have also performed a few runs for which the
interaction is turned off. Runs of type $C$ involve two shells of
fragments.

\vspace*{2mm} \noindent
\begin{table*}[hbt!]
\begin{center}
\begin{tabular}{ccccc}
\hline \hline Run & $M_s/M$ & $R_s/R$ & $t_{fall}/t_{cr}$&Notes\\
\hline $A_1$ & 0.1     & 0.1   &  48.125 &        \\
$A_6$ & 0.05    & 0.1   &  110.0  &\\
$A_8$ & 0.05    & 0.05  & 73.15 & $r_{min}=0.05~R$\\ $A_9$ & 0.05
& 0.025 &  60.5 & $r_{min}=0.022~R$\\ \hline
\end{tabular}
\end{center}
\caption{\small{Single satellite, $r_s (0)= R$.}}
\label{tab:tab2a}
\end{table*}







\begin{table*}[hbt!]
\begin{center}
\begin{tabular}{cccccc}
\hline
\hline
Run   & $N_f$ & $M_f/M$  & $r_{shell}(0)/R$&$t_{fin}/t_{cr}$&
Notes\\ \hline $B_1$ & 100   & 0.001  & 1 &$220$& case I\\ $B_2$ &
100   & 0.001  & 0.2 &$110$  &  case I\\ $B_3$& 100   & 0.001  &
0.4        &$110$  & case I\\ $B_4$ & 100   &$5 \cdot 10^{-4}$ & 1
&$220$  & case I\\ $B_5$ & 100 &$5 \cdot 10^{-4}$ & 0.2 &$110$ &
case I\\ $B_6$ & 100 & $5 \cdot 10^{-4}$ & 0.4 &$110$  & case I\\
$BS_1$ & 20 &$0.005$ & 0.3     &220 & case II
\\ $BS_2$ & 100 & $0.001$ & 0.3 &220 & case II \\ $BS_3$ & 20
& $0.005$ & 1.      &110 & case II\\ $BS_4$ & 100 & $0.001$ & 1.
&110 & case II \\ $BT_1$& 20 & $0.005$ &  0.3 &220 & case III\\
$BT_2$ & 100 & $0.001$ & 0.3      &220 & case III\\ $BT_3$ & 20 &
$0.005$ & 0.7      &110 & case III\\ $BT_4$ & 100 & $0.001$ & 0.7
     &110 & case III\\ $C_1$& 200   &  $5 \cdot 10^{-4}$ &  1, 0.2    & $220$  & case I\\
$C_2$& 200   &   $5 \cdot 10^{-4}$& 1, 0.4    & $220$  & case I\\
\hline
\end{tabular}
\end{center}
\caption{\small{Simulations with shells of fragments. }}
\label{tab:tab2b}
\end{table*}





\subsection{The case of one sinking satellite}
\label{sect:single_result}
We have studied the energy transfer between the galaxy and the
sinking satellite and the overall energy balance, using the total
energy conservation as a diagnostic of the code accuracy. In
particular we have considered the quantity $\Delta E_{sat} =
K_{sat} + E_{int} - K_{sat}(t = 0) - E_{int}(t = 0)$. For example,
for run $A_1$ (characterized by $M_s/M = 0.1$, $R_s/R = 0.1$,
$t_{fall}/t_{cr} = 48.125$) the energy lost by the satellite is
initially very small (on the order of $0.02$, in the units
following the conventions described in Sect.~\ref{sect:units}, at time $t
\approx 38.5~t_{cr}$), and then rapidly rises (to $0.075$ at time
$t \approx 49.5~t_{cr}$); during the run the total energy is
conserved to better than $0.002$. Similarly, we have considered
the angular momentum balance. Here the relatively large amount of
angular momentum already associated with the galaxy right after
the beginning of the simulation ($J_{gal} \approx 0.0054$ at $t
\approx 5.5~t_{cr}$) is not yet due to the scattering of star
orbits by the sinking satellite, but is mostly related to the
overall orbital motion of the galaxy in the inertial frame of
reference of the center of mass; eventually, the satellite is
dragged in, down to the center, and loses completely its angular
momentum, which is acquired by the galaxy.

During the process, following the steps indicated by Bontekoe \&
van Albada (1987), we can ``measure" the Coulomb logarithm. The
measurement of $\ln{\lambda}$ consists of the following steps: (i)
Determine the velocity $\hat{v}_s$ of the satellite with respect
to the local streaming motion of the stars (the streaming motion
is determined as a mean velocity of the stars in the ``vicinity''
of the satellite); (ii) Measure the energy $E_{sat}$ of the
satellite at many instants, then determine $dE_{sat}/dt;$ (iii)
Determine the density $\rho_G$ of the galaxy at the position of
the satellite by linear interpolation of densities between the 8
nearest mesh corners; (iv) Calculate $F(\hat{v}_s)$, the fraction
of particles with velocity smaller than $\hat{v}_s$. Finally the
Coulomb logarithm is estimated from the relation $ 4\pi G^2
\ln\lambda = - v_s (dE_{sat}/d t)/(M_s^2\rho_G F(\hat{v}_s))$.

\begin{figure*}[hbt!]
\centerline{\epsfig{file=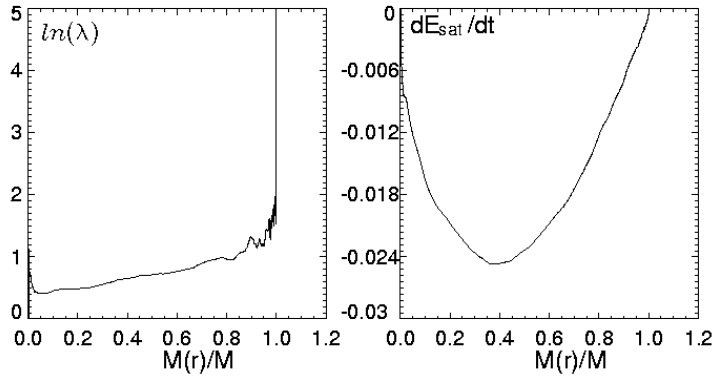,height=5.5cm,angle=0} }
\caption{\small{Numerical
measurement of the Coulomb logarithm (left)
           and of the satellite energy loss by dynamical friction (right)
           for run $A_1$ (characterized by $M_s/M = 0.1$, $R_s/R = 0.1$,
$t_{fall}/t_{cr} = 48.125$). The quantities are plotted against a
Lagrangian
           radial coordinate in the same format as in the paper by Bontekoe
           \& van Albada (1987).}}
\label{fig:fig1}
\end{figure*}

\begin{figure*}[hbt!]
\centerline{
\epsfig{file=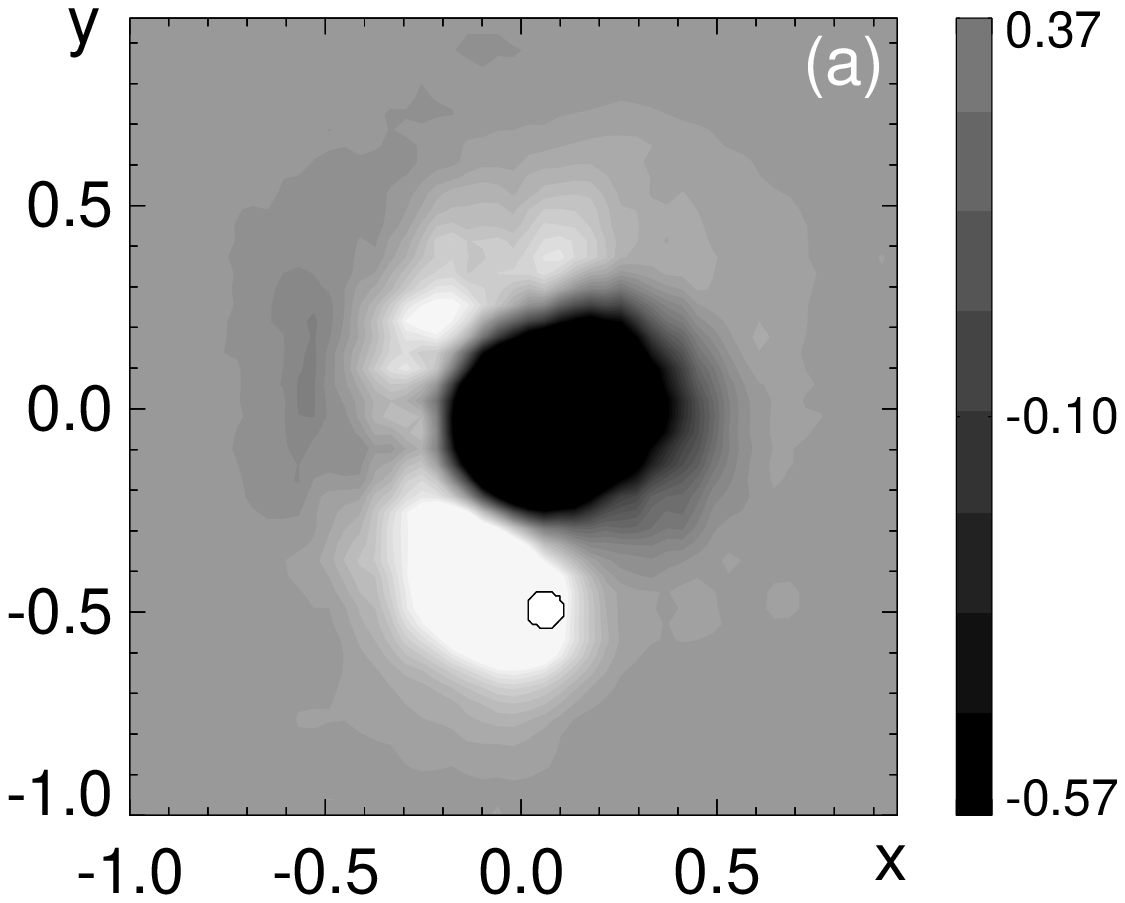,height=5.5cm,angle=0}
\epsfig{file=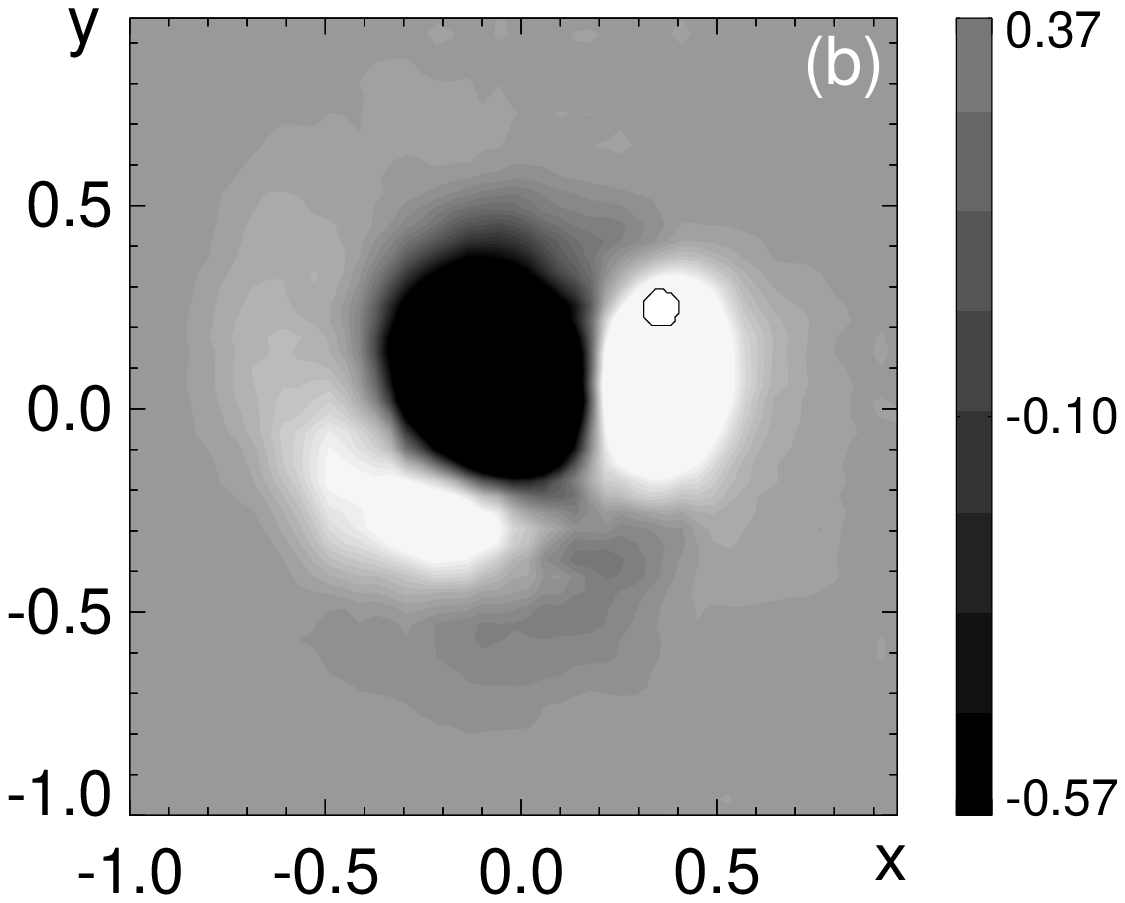,height=5.5cm,angle=0}
}
\centerline{
\epsfig{file=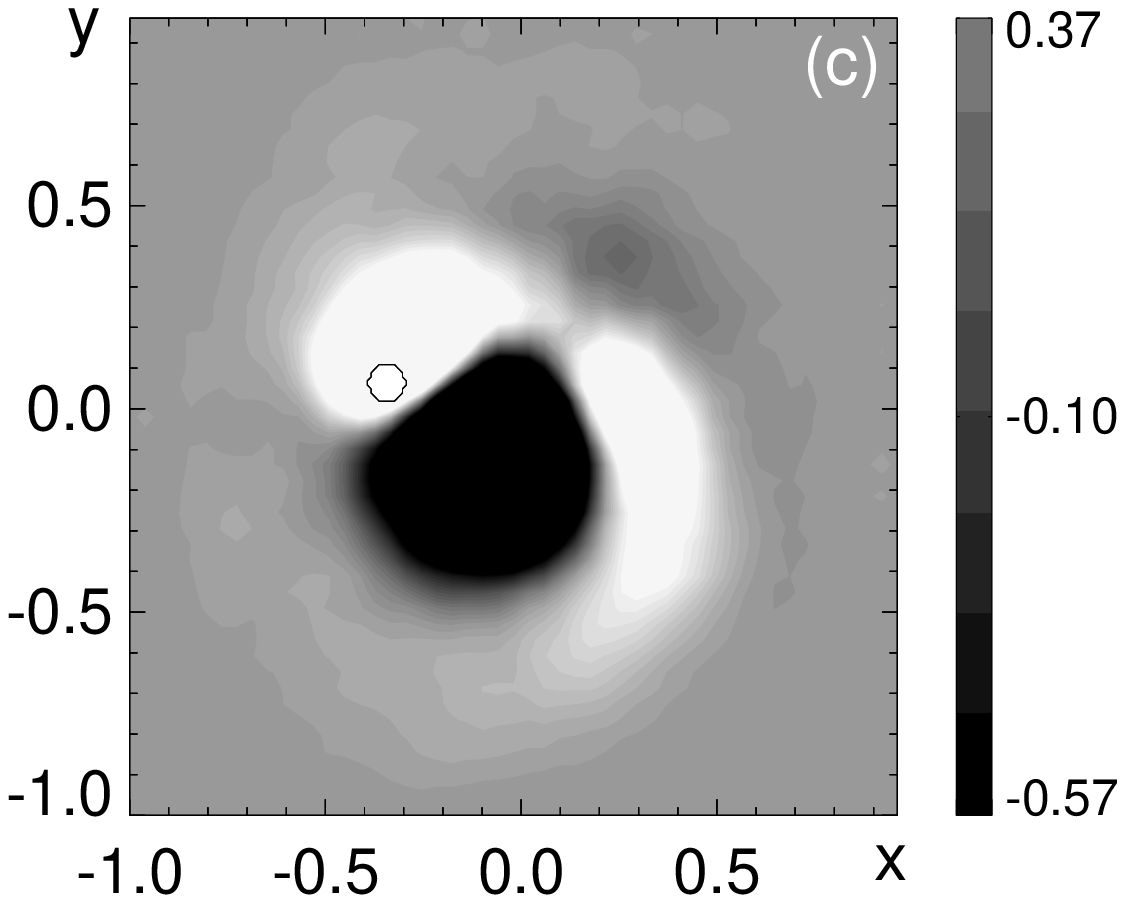,height=5.5cm,angle=0}
\epsfig{file=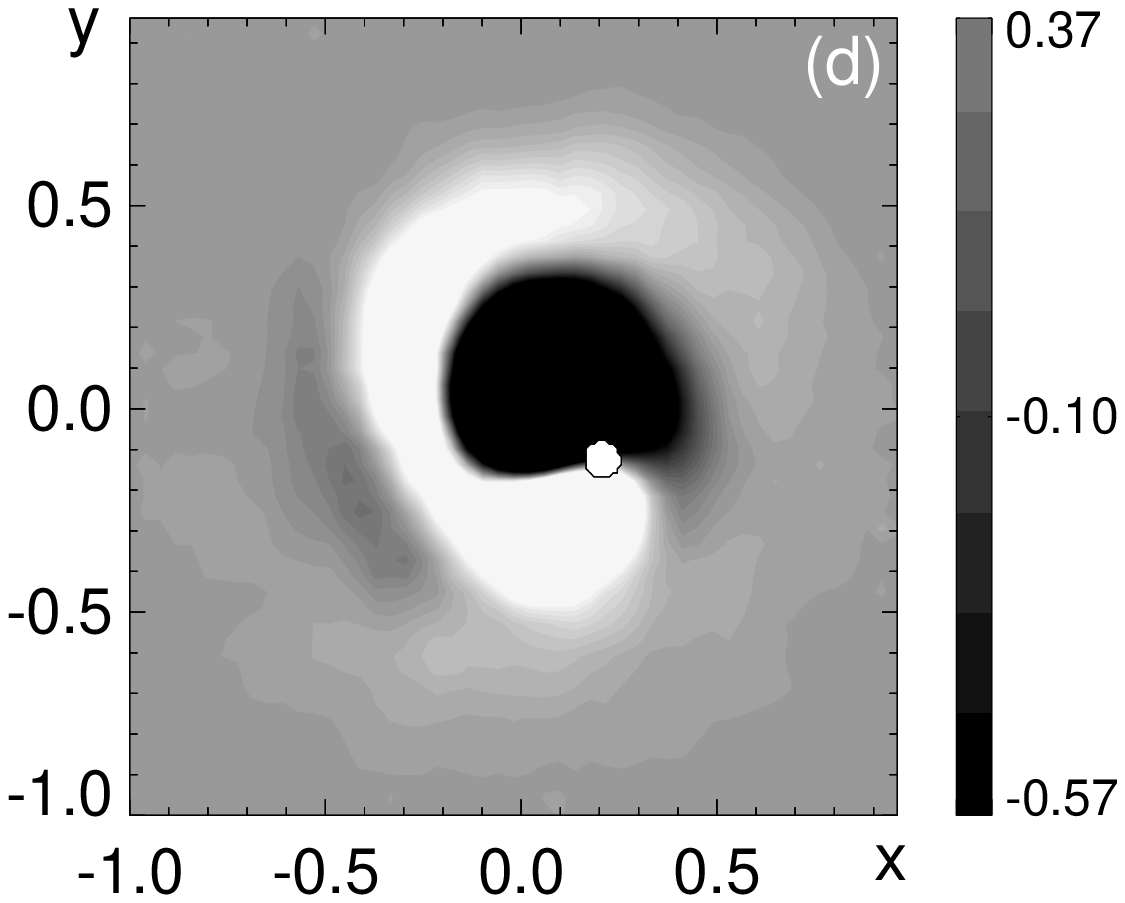,height=5.5cm,angle=0}
}
\centerline{
\epsfig{file=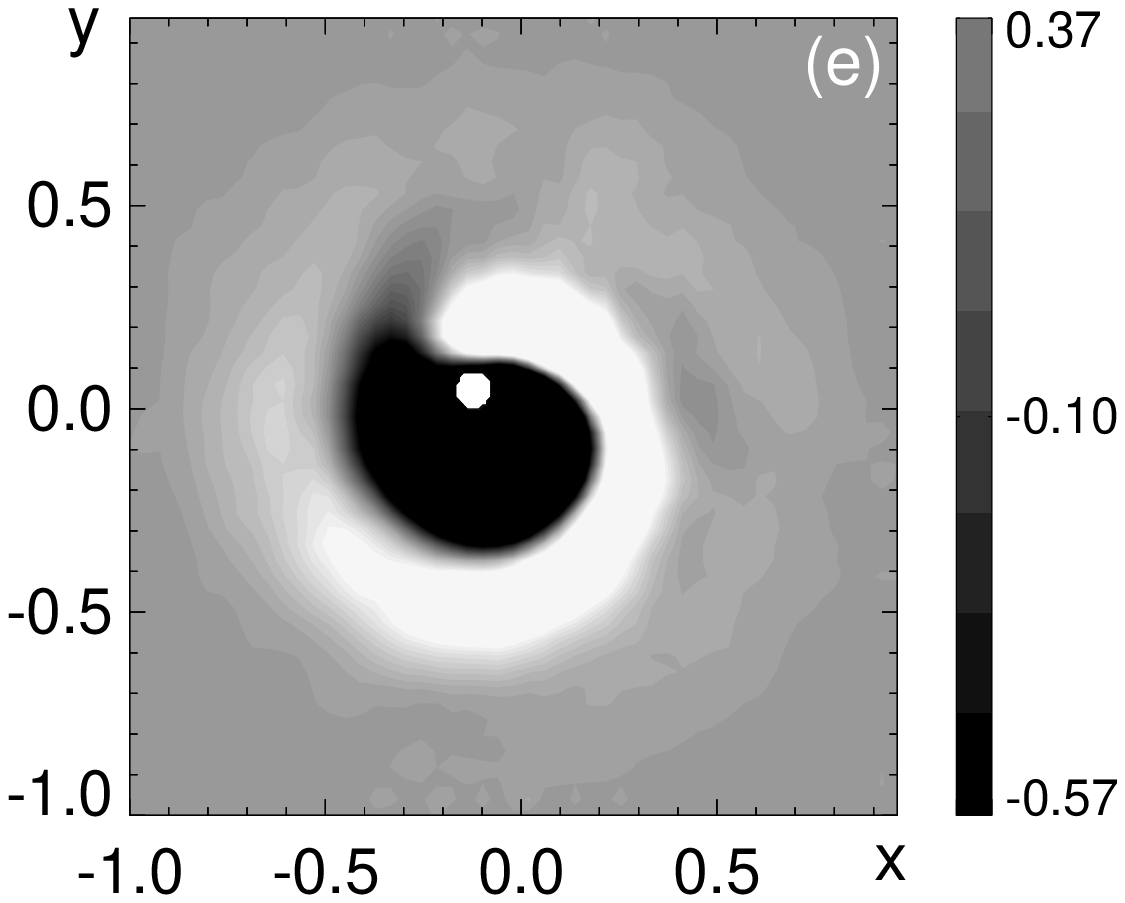,height=5.5cm,angle=0}
\epsfig{file=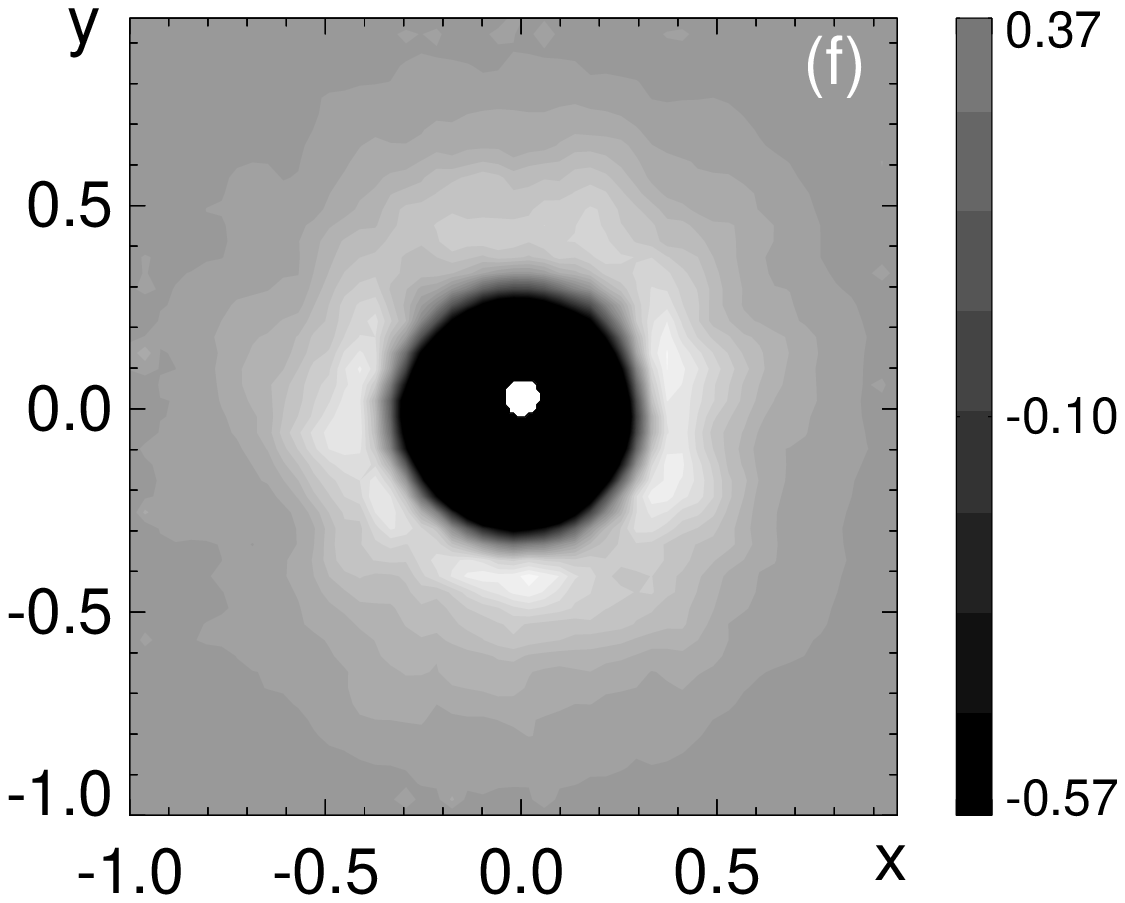,height=5.5cm,angle=0}
}
 \caption{\small{ Six frames describing the equatorial density response
$\rho_1 = \rho - \rho_0$ of the galaxy to the infalling satellite
for run $A_1$. The density wakes are portrayed at time $t/t_{cr}
=$ (a) 38.5, (b) 39.875, (c) 41.25, (d) 42.625, (e) 44, and (f)
53.625. The white circle denotes the location of the satellite.}}
\label{fig:fig2}
\end{figure*}

\begin{figure*}[hbt!]
\centerline{ \epsfig{file=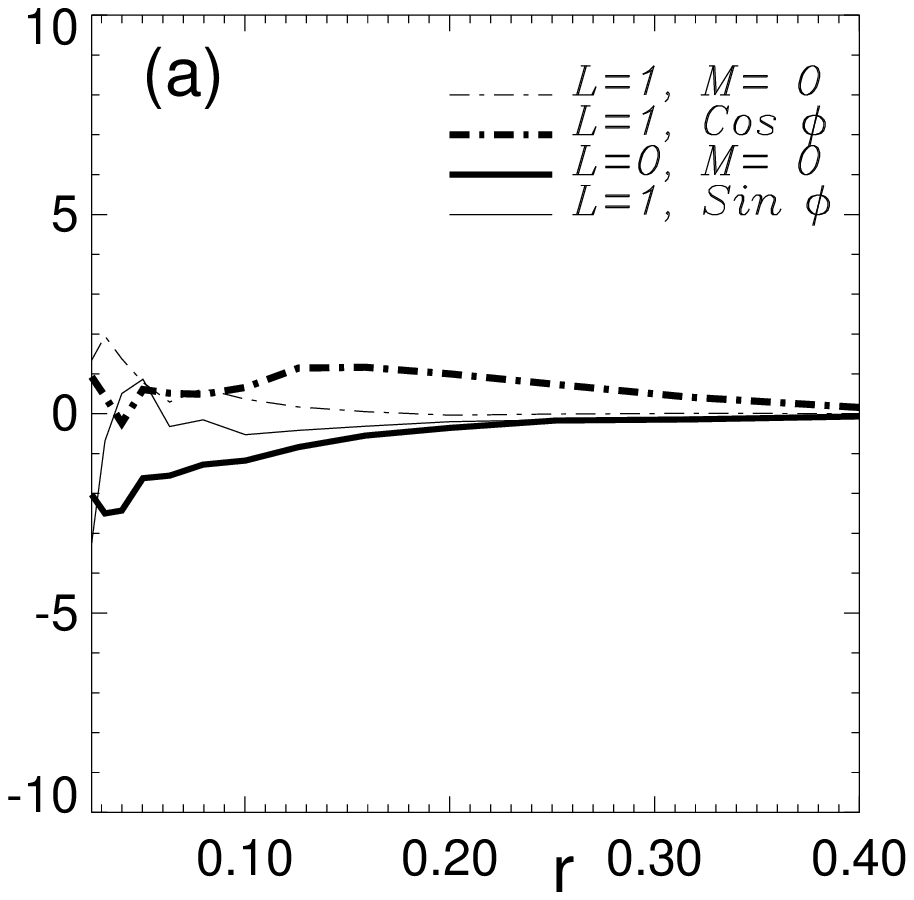,height=5.5cm,angle=0}
\epsfig{file=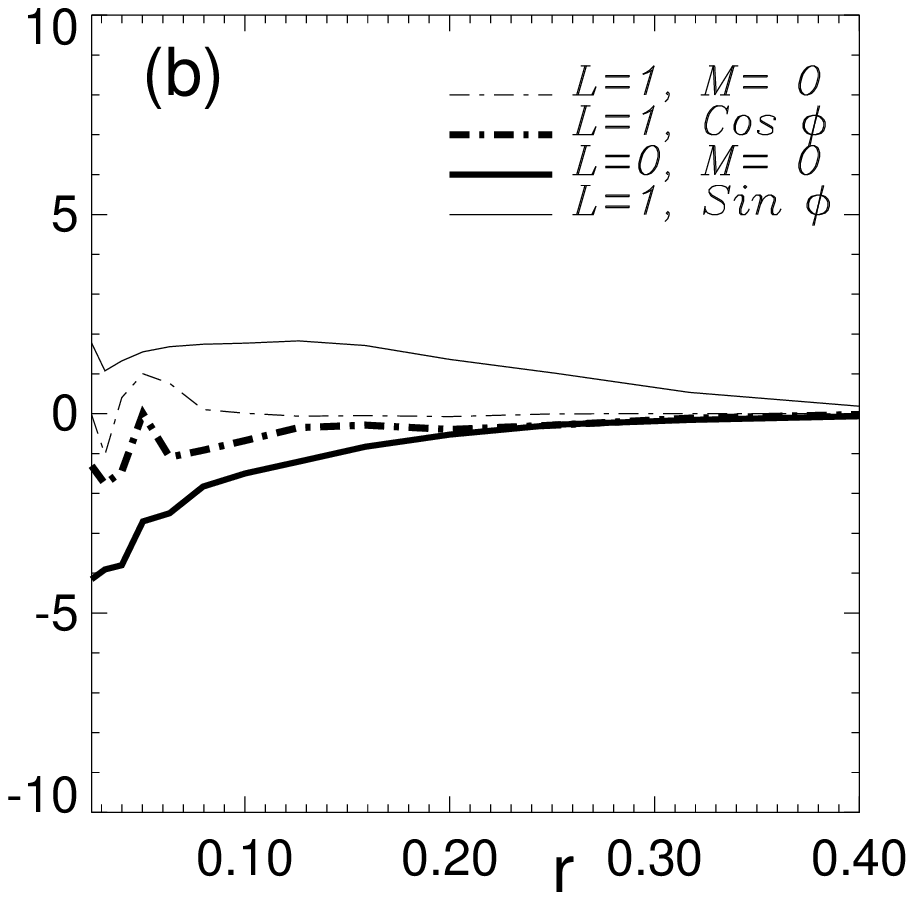,height=5.5cm,angle=0} } \centerline{
\epsfig{file=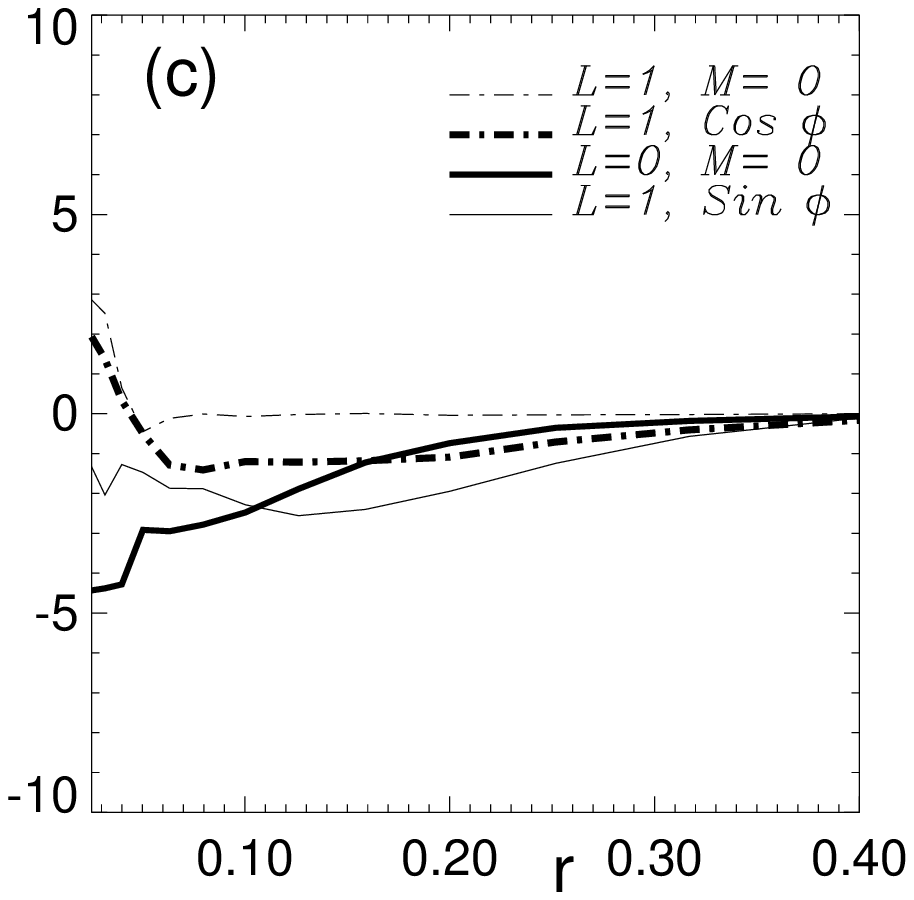,height=5.5cm,angle=0}
\epsfig{file=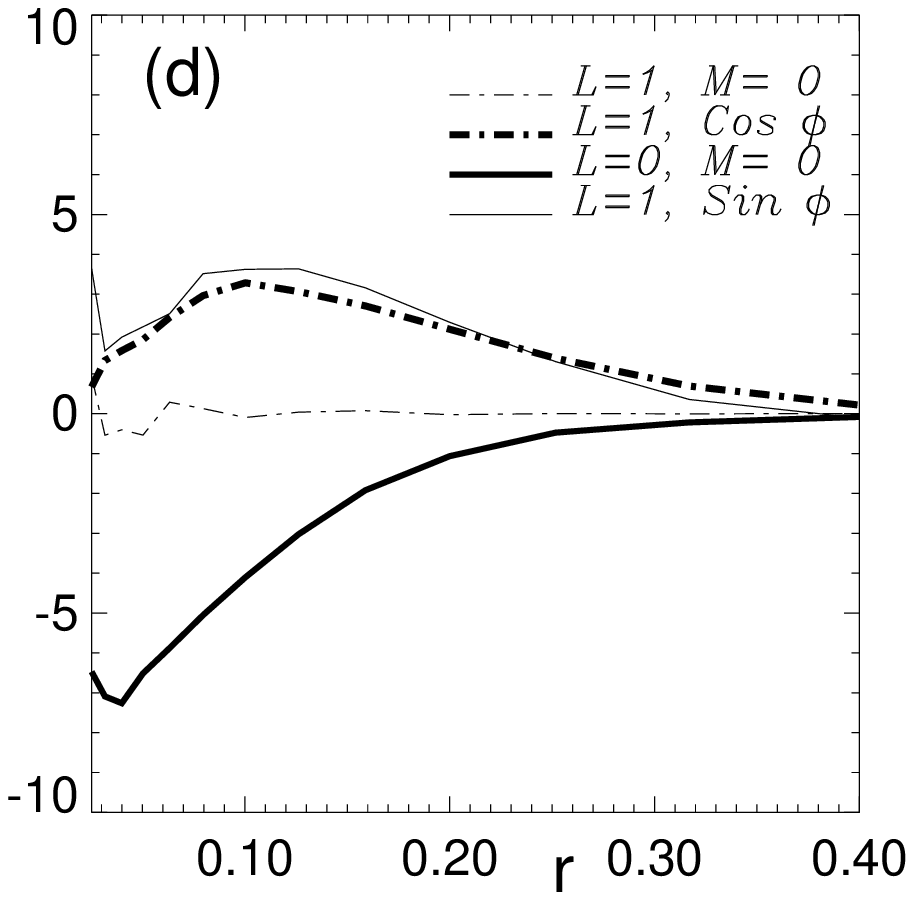,height=5.5cm,angle=0} } \centerline{
\epsfig{file=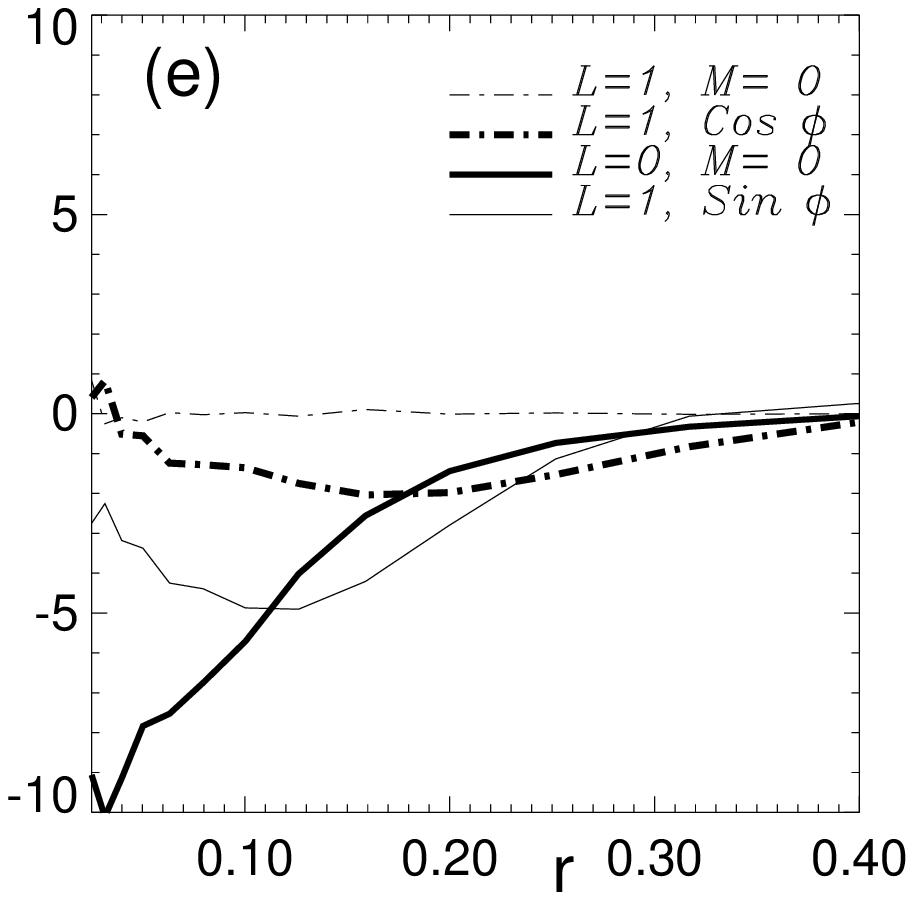,height=5.5cm,angle=0}
\epsfig{file=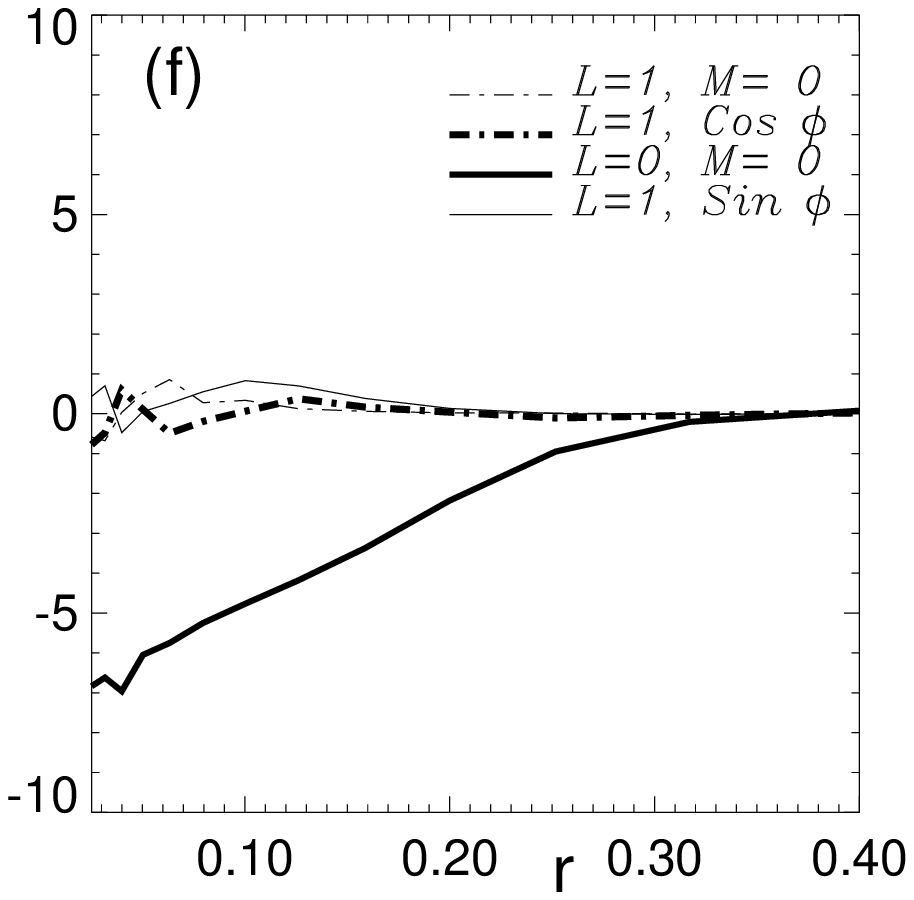,height=5.5cm,angle=0} }  \caption{\small
For each frame of Fig.~\ref{fig:fig2} we illustrate in detail the
monopole $(l=0)$ and the dipole $(l=1)$ contributions to the
density wake in the galaxy (the various curves represent the
coefficients $A_{10}$, $A_{11}$, $A_{00}$, and $B_{11}$ defined in
the text; see Sect.~\ref{sect:pressure}).}
\label{fig:fig3}
\end{figure*}

\begin{figure*}[hbt!]
\centerline{ \epsfig{file=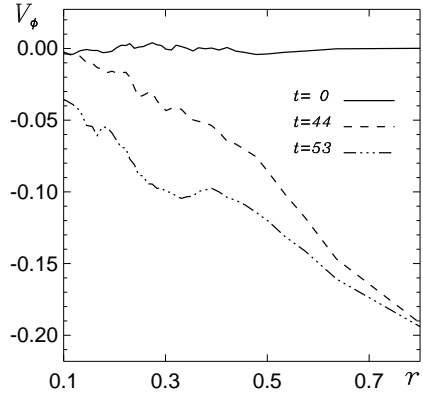,height=5.5cm,angle=0}}
\caption{\small For run $A_1$ we plot the mean azimuthal velocity
of the galaxy measured on the equatorial plane at three different
times; the rotation induced by the angular momentum exchange with
the infalling satellite is approximately that of a solid body. As
usual, time is measured in units of $t_{cr}$.} \label{fig:fig4}
\end{figure*}

\begin{figure*}[hbt!]
\centerline{
\epsfig{file=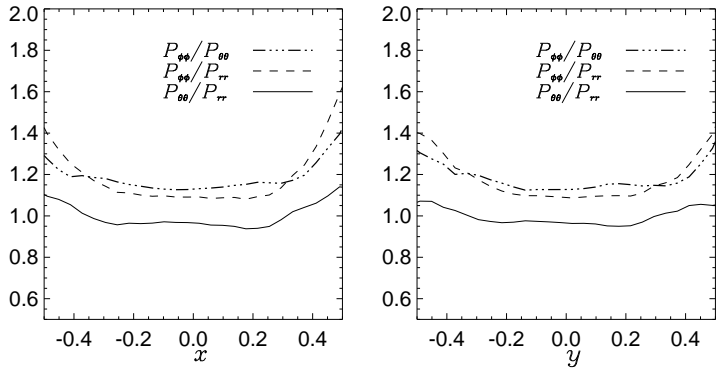,height=5.5cm,angle=0}} \caption{\small
For run $A_1$ we plot two othogonal equatorial cuts of the anisotropy
distribution generated in the galaxy by the infalling satellite at
time $t = 53.625~t_{cr}$, after the satellite has reached the
center.}
\label{fig:fig5}
\end{figure*}

\begin{figure}[hbt!]
\centerline{
\epsfig{file=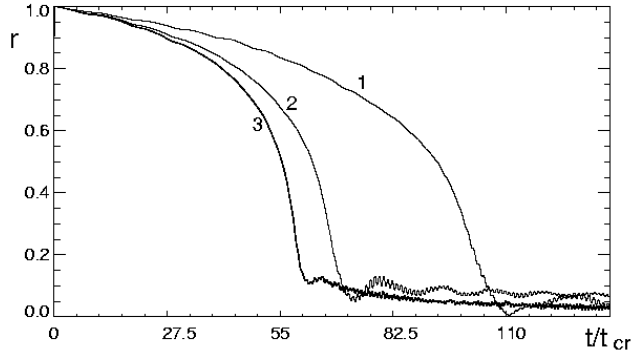,height=5.5cm,angle=0}}
 \caption{\small The curves
illustrate the orbital decay process for the satellite for three
different runs ($A_6$, $A_8$, and $A_9$, labeled as 1, 2, and 3
respectively) characterized by different spatial size of the
infalling satellite ($R_s = 0.1,~0.05,~0.025~R$; $M_s/M = 0.05$).}
\label{fig:fig6}
\end{figure}

\begin{figure*}[hbt!]
\centerline{ \epsfig{file=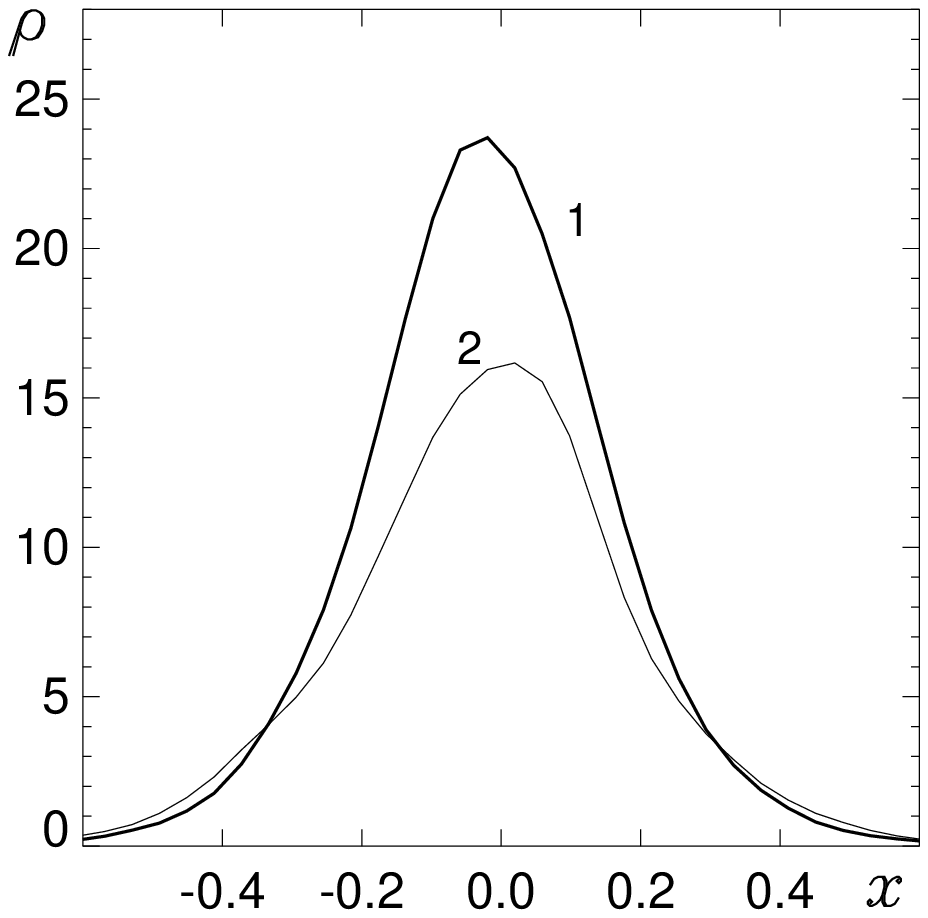,height=4.8cm,angle=0}
\psfig{file=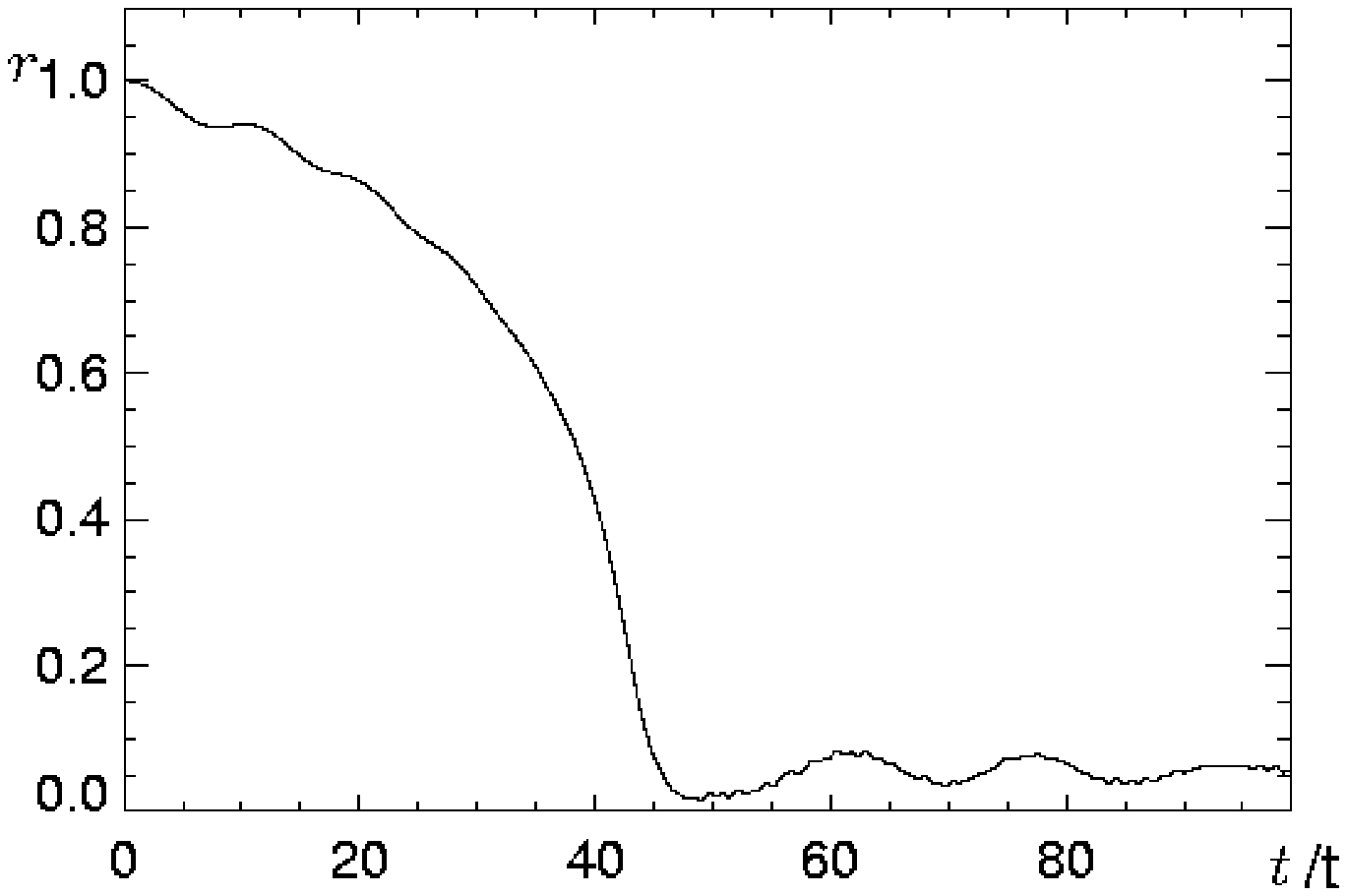,height=4.8cm,angle=0} }  \caption{\small
Left: The density distribution change induced in the galaxy by the
infall of the satellite (run $A_1$). The thick upper curve (1)
represents the initial distribution, while the thin lower curve
(2) is associated with the final state. Right: Orbital decay for
run $A_1$ (cf. Fig.~\ref{fig:fig2}) as in Fig.~\ref{fig:fig6}.}
\label{fig:fig7}
\end{figure*}

The main results that we have obtained by examining the
experiments for the case of a single sinking satellite are the
following:

\begin{itemize}
  \item Qualitatively, the general conclusions derived by Bontekoe \& van Albada
  (1987) have been checked to hold. However, we have found that,
  while the classical formula of dynamical friction can be used
  to fit the orbital decay of the satellite over a significant
  time interval, the determined value of the Coulomb logarithm changes
  significantly in the course of the simulation (see Fig. \ref{fig:fig1}); during the
  initial transient and the final
  part of the orbital evolution, when the satellite reaches the
  central regions, the change in the nominal value of the Coulomb logarithm can be
  dramatic, as already noticed by Bontekoe \& van Albada (1987).

  \item The process of dynamical friction has been checked to
  operate via the generation of wakes, of which the dipolar $l =
  1$ component appears to play a dominant role (see Figs. \ref{fig:fig2} and
  \ref{fig:fig3}). This generally agrees with the scenario outlined by Weinberg
  (1989); see also Sect.~\ref{sect:theory}.

  \item As a result of the scattering of stars due to the sinking
  satellite, the galaxy acquires not only some angular momentum (the induced systematic motions are approximately those of
  solid body rotation, see Fig. \ref{fig:fig4}), but also a significantly
  anisotropic pressure tensor. Figure \ref{fig:fig5}, relative to run $A_1$,
  suggests that the final configuration reached is reasonably well
  characterized by $p_{rr} \sim p_{\theta\theta}$, while the
  toroidal pressure $p_{\varphi\varphi}$ is significantly larger
  even in the central regions. As a result, except for the outer
  parts, the evolved distribution function of the galaxy may turn
  out to be accessible to models with $f = f(E, J_z)$.

  \item In some runs, for which the diameter of the satellite
  is reduced, the satellite turns out to fall inwards faster, but
  to be unable to reach the center of the galaxy (see Fig.~\ref{fig:fig6}).
  We have checked that this effect does not depend on
  the number N by running a simulation of type $A_9$ with N up to
  $2 \cdot 10^6$ (in such a way that the number of simulation
  particles in the region inside the radius of minimum approach increases
  from about 600 to about 2500; correspondingly, the observed
  radius of minimum approach changes by less than 1 \%). We
  should recall that simulations with a code of the type we
  are using, with a finite expansion in spherical harmonics,
  cannot deal realistically with spatially small satellites (see
  comment at the end of Sect.~\ref{sect:single_star}).
  Still, a ``saturation" of
  dynamical friction of the kind noted in our simulations is likely
  to be real, because the modeling problem associated with the
  spherical harmonic expansion is less severe for the late phases
  of the simulation, when the satellite is close to the galaxy
  center. The ultimate fate of a single sinking object might be
  related to other interesting issues, such as those met in the
  discussion of the formation of massive black hole binaries (e.g.,
  see Quinlan \& Hernquist \cite{qui97}, Milosavljevich \& Merritt
  \cite{mil01}), which go well beyond the scope of the present
  paper (that addresses the slow evolution of a galaxy on the large
  scale as a result of the capture of a large number of small
  objects); what we have noted here is just meant to say
  that on this specific feature we confirm the results of previous
  investigations (in particular,
  those of Bontekoe \& van Albada 1987) carried out with a much smaller
  number of particles.

  \item Finally we note that the overall density distribution of
  the galaxy, at the end of the simulation, turns out to be {\it
  less concentrated} with respect to the initial distribution (see
  Fig.~\ref{fig:fig7}). In passing, we note that a figure such as
  Fig.~\ref{fig:fig7} naively suggests
  that the total mass is not conserved, but we have checked that in this
  respect this is only a false impression left by our way of representing
  the density profile; in particular, one should recall that the weight of the
  volume element rapidly increases with radius.
\end{itemize}

\vspace*{-10mm}


\subsection{The case of many fragments}
\label{sect:many_result}
\subsubsection{Case I}
\label{sect:many_result1}
For various runs of type $B$ and $C$ we have checked that the
total energy conservation generally holds, over a period of
$\approx 100~t_{cr}$, to better than one part over 500. Yet, it is
difficult to estimate the impact of dynamical friction on the
galaxy evolution, because secular effects are masked by effects
related to the initial lack of equilibrium. From this class of
numerical experiments we have obtained the following results:

\begin{itemize}

  \item We have observed a process of shell {\it diffusion}, which
  is a collective effect qualitatively different from what might
  have been imagined by simply superposing the effects of orbital
  decay of the individual fragments.
  The phenomenon
  persists, presumably via interaction with the wakes induced by the fragments
  in the galaxy, even when the direct fragment-fragment interaction is turned
  off. Sharp coherent oscillations are noted at early times;
  they are recognized as
  epicyclic oscillations generated by our choice of initial
  velocities for the fragments.
  We have also performed tests on the process of shell diffusion and oscillations
  on the basis of 400 fragments initially located close to $r = 0.6~R$.

  \item We have noted systematic changes of the galaxy density distribution. The
  general trend emerging from the simulations is that significant
  {\it peaking} of the density distribution may occur, {\it provided the
  initial radius of one shell of fragments is well inside the
  galaxy}.
  However, much of the change is
  apparently associated with the initial lack of equilibrium.

  \item The process of density profile modification is not accompanied by the
  development of significant pressure anisotropy.

\end{itemize}

In conclusion, simulations of Case I turn out to be of little
direct use for the main objectives of our project. However, {\it
they are extremely useful and instructive}, because they
demonstrate that, in order to properly single out slow relaxation
processes, one has to be very careful about the choice of initial
conditions and, in particular, because they provide concrete
examples of some misleading effects that can occur as a result of
the use of improper models.

\subsubsection{Case II}
\label{sect:many_result2}
\begin{figure*}[hbt!]
\leftline{ \epsfig{file=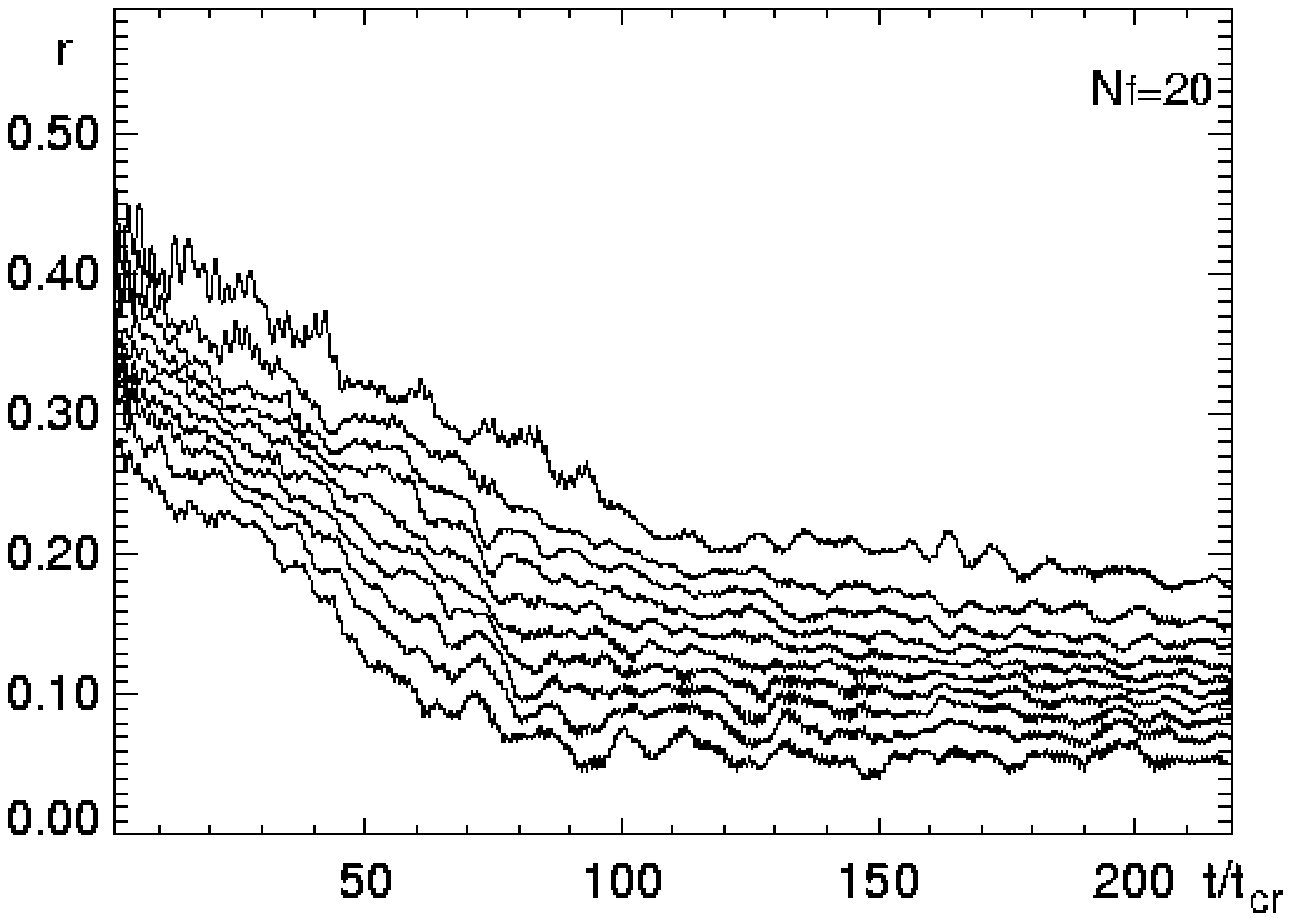,height=4.9cm,angle=0}
\epsfig{file=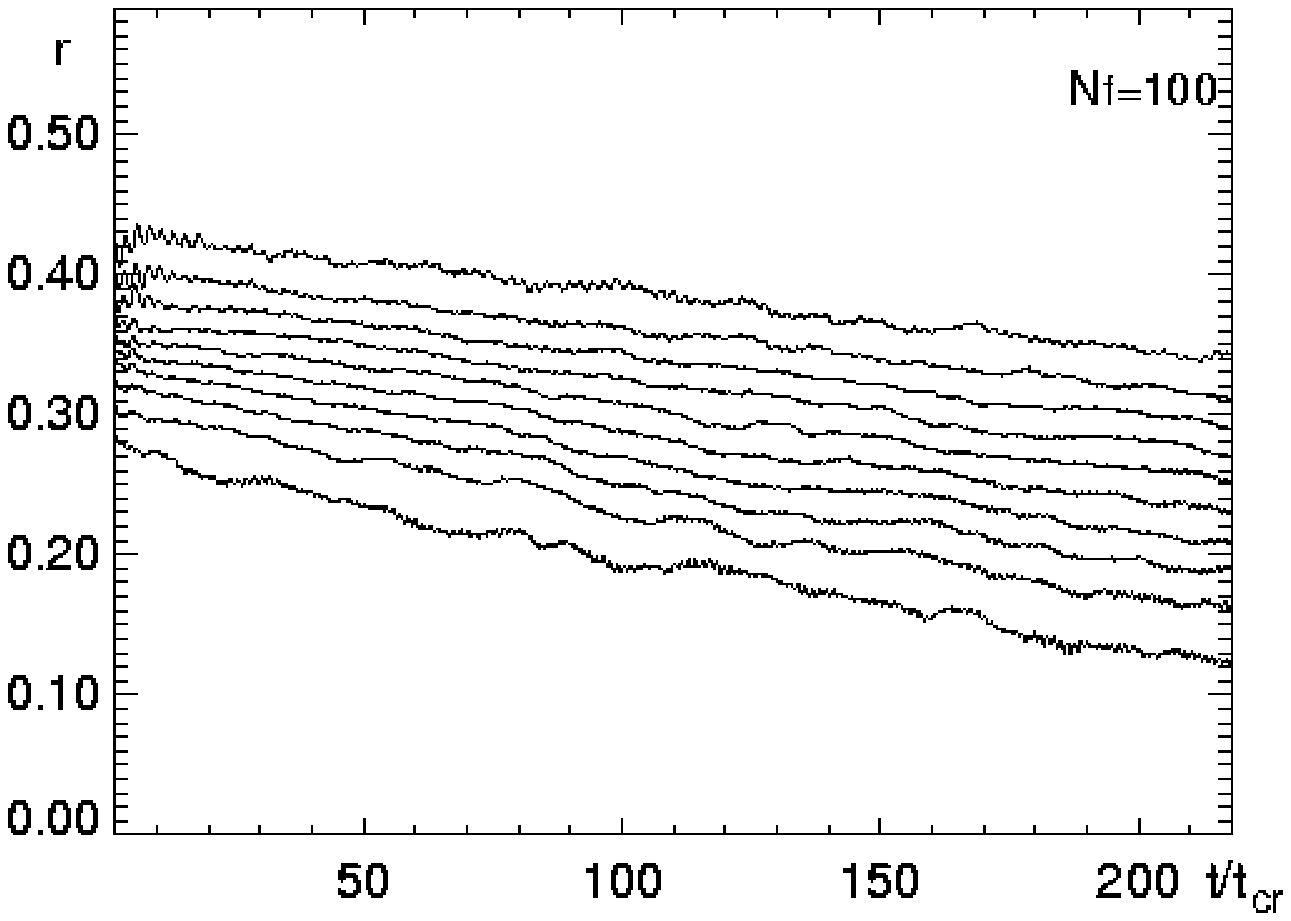,height=4.9cm,angle=0}}
\leftline{\epsfig{file=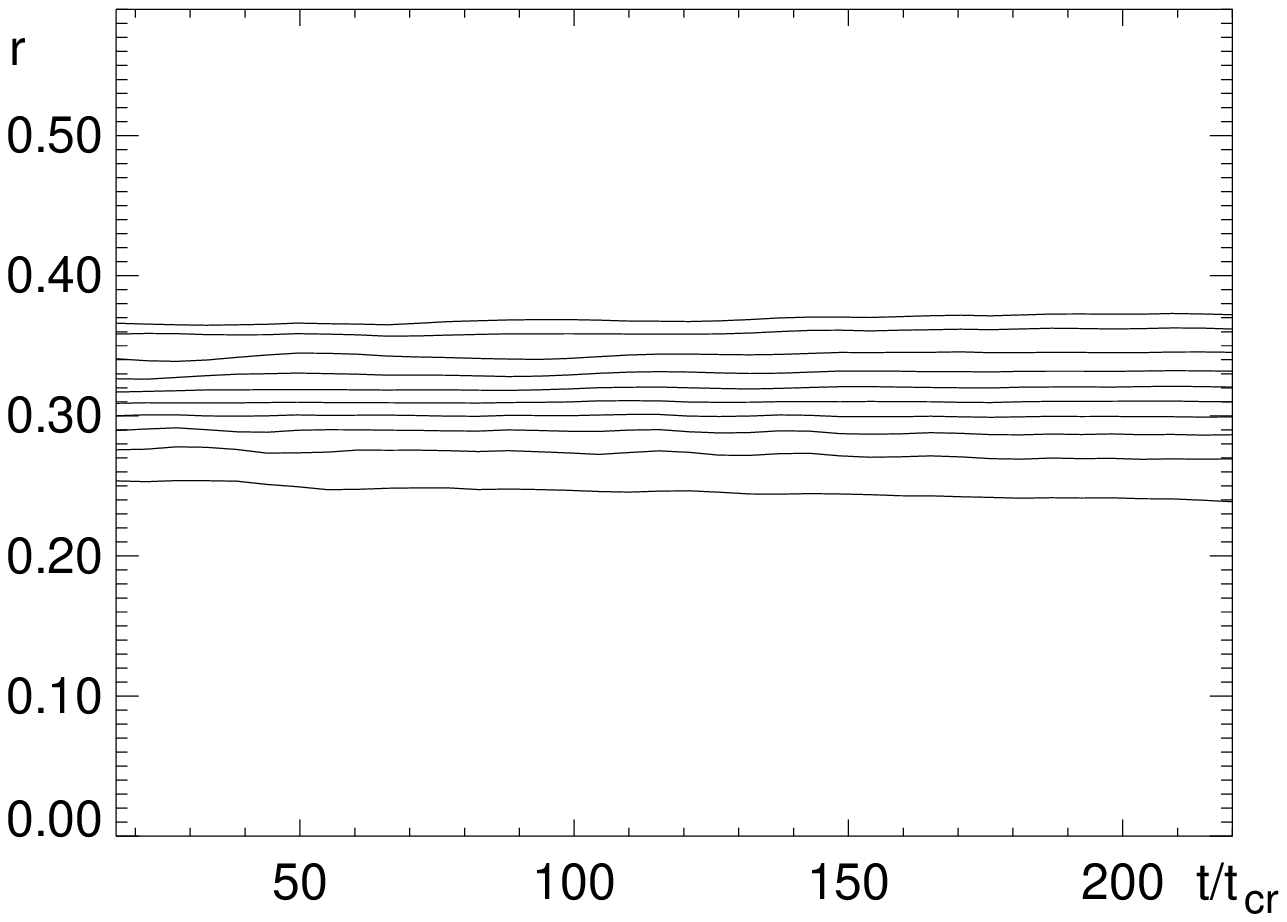,height=4.9cm,angle=0}}
\vspace*{-51mm}
\rightline{
\epsfig{file=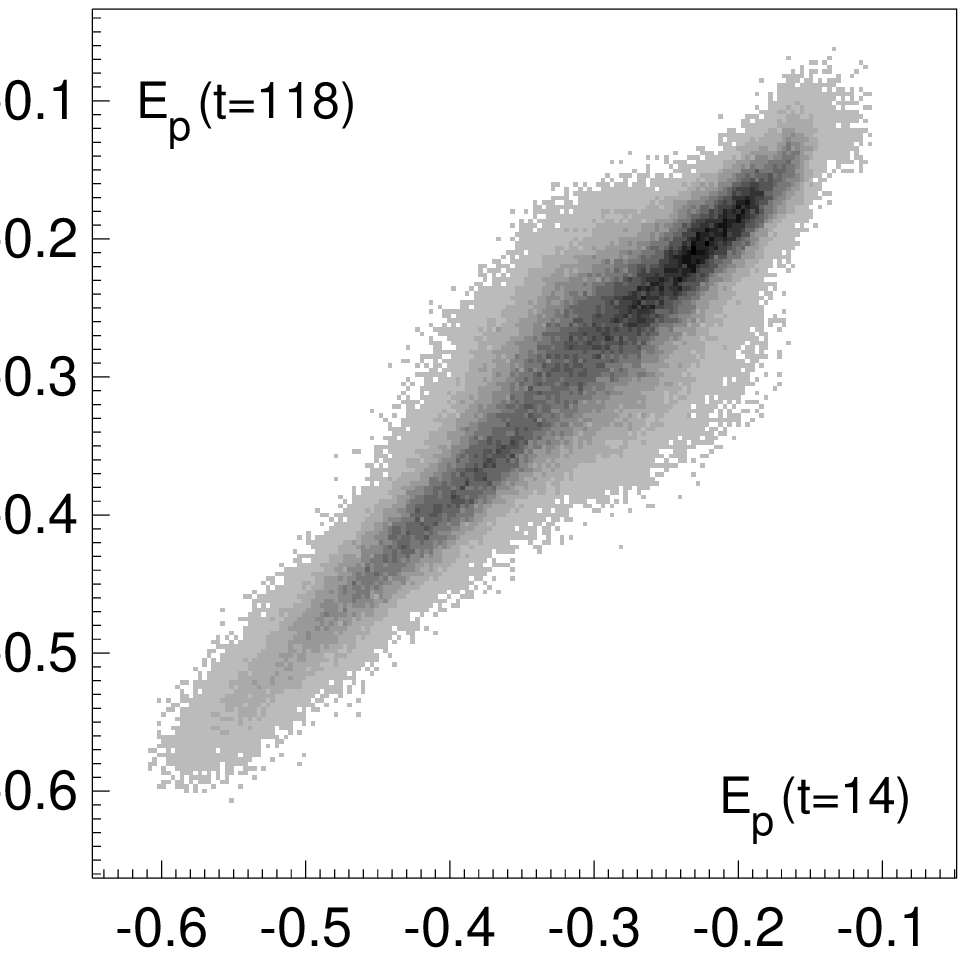,height=4.9cm,angle=0} } \caption{\small
The fall of a shell of fragments towards the galaxy center, for
the case of $N_f = 20$ (left frame on the top, run $BS_1$) and for
$N_f = 100$ (right frame on the top, run $BS_2$). For comparison,
the left frame at the bottom describes the time evolution of a
collisionless shell (a run with $N_f = 5 \cdot 10^4$). Displayed
lines represent 5, 15, 25, 35, 45, 55, 65, 75, 85, and 95 \% of
the shell mass. The right frame at the bottom displays, for run
$BS_2$, the correlation between single particle energy at time $t
= 14~ t_{cr}$ and energy at time $t = 118 ~t_{cr}$, showing the
effects of scattering associated with dynamical friction; units
for energy are those indicated in Sect.~\ref{sect:units}.}
\label{fig:fig13}
\end{figure*}

\begin{figure*}[hbt!]
\leftline{ \epsfig{file=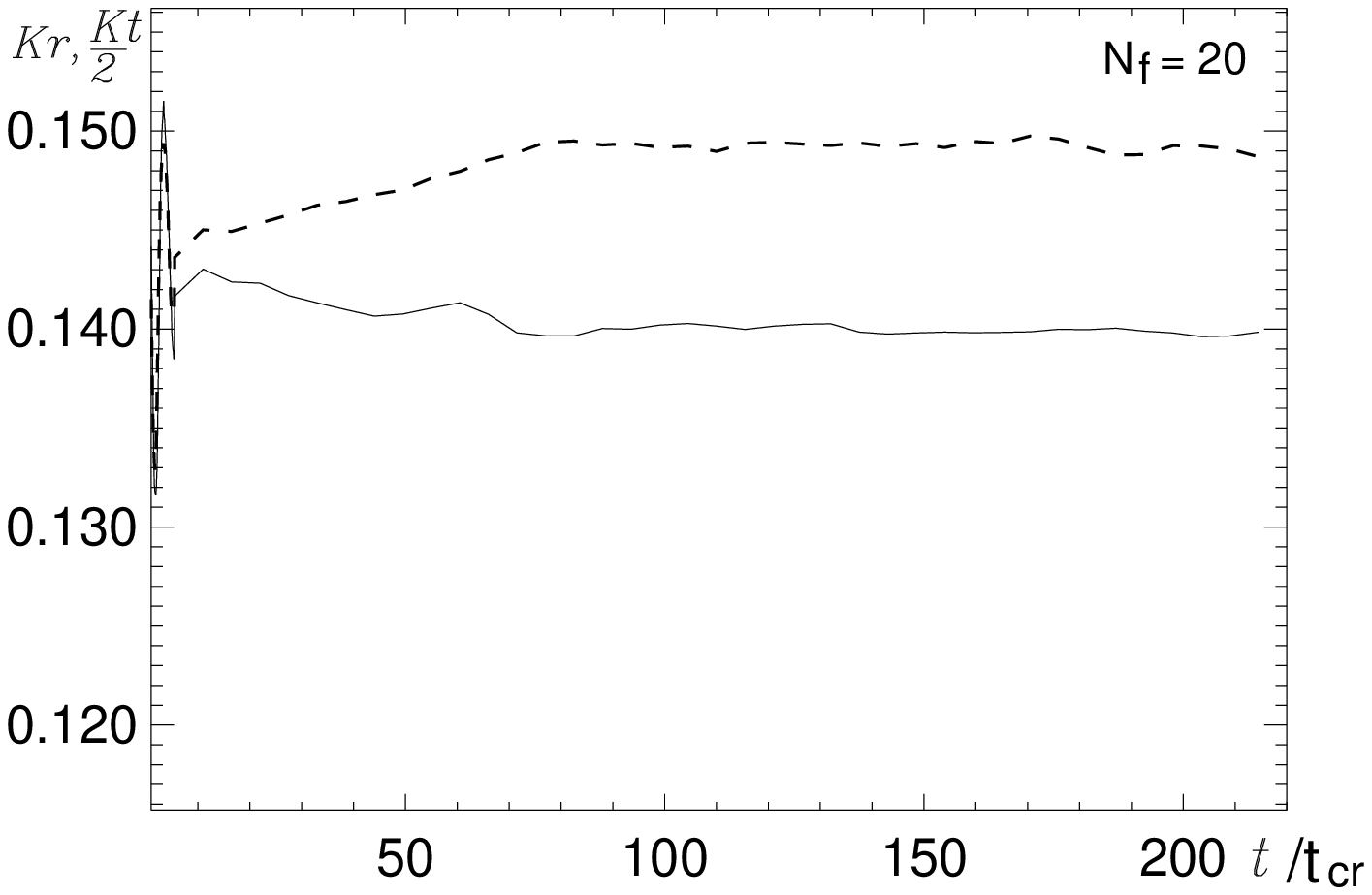,height=4.9cm,angle=0}
\epsfig{file=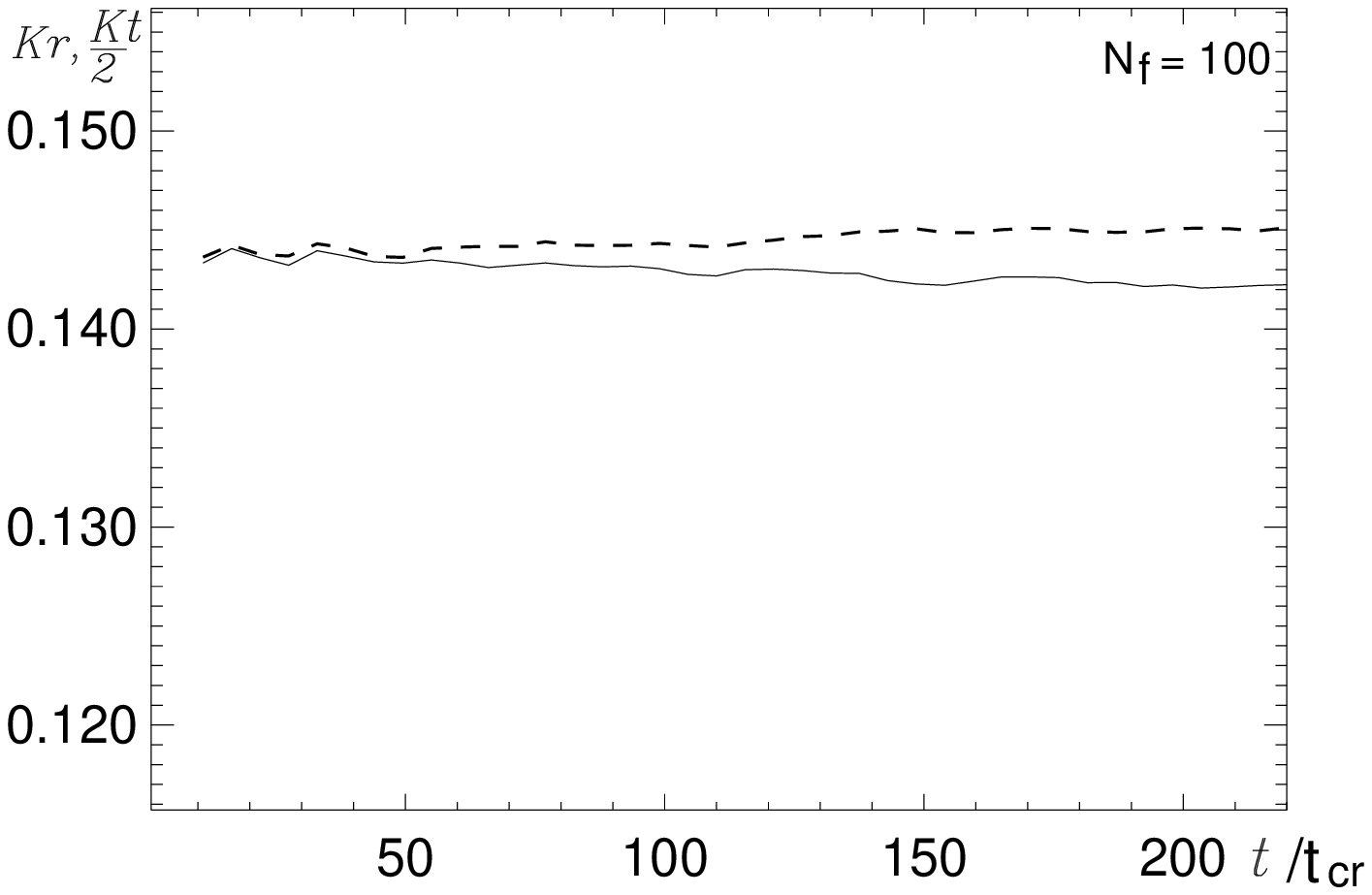,height=4.9cm,angle=0}} \caption{\small
The development of pressure anisotropy in the galaxy as a result
of the interaction with a shell of fragments dragged in towards
the galaxy center, for the case of $N_f=20$ (left frame, run
$BS_1$) and for $N_f=100$ (right frame, run $BS_2$). The dashed
curves represent the evolving value of $K_T/2$, where $K_T$ is the
total kinetic energy associated with the star motions in the
tangential directions; the solid curve represents the evolving
value of $K_r$, the total kinetic energy associated with the star
motions in the radial direction.} \label{fig:fig14}
\end{figure*}

\begin{figure*}[hbt!]
\centerline{ \epsfig{file=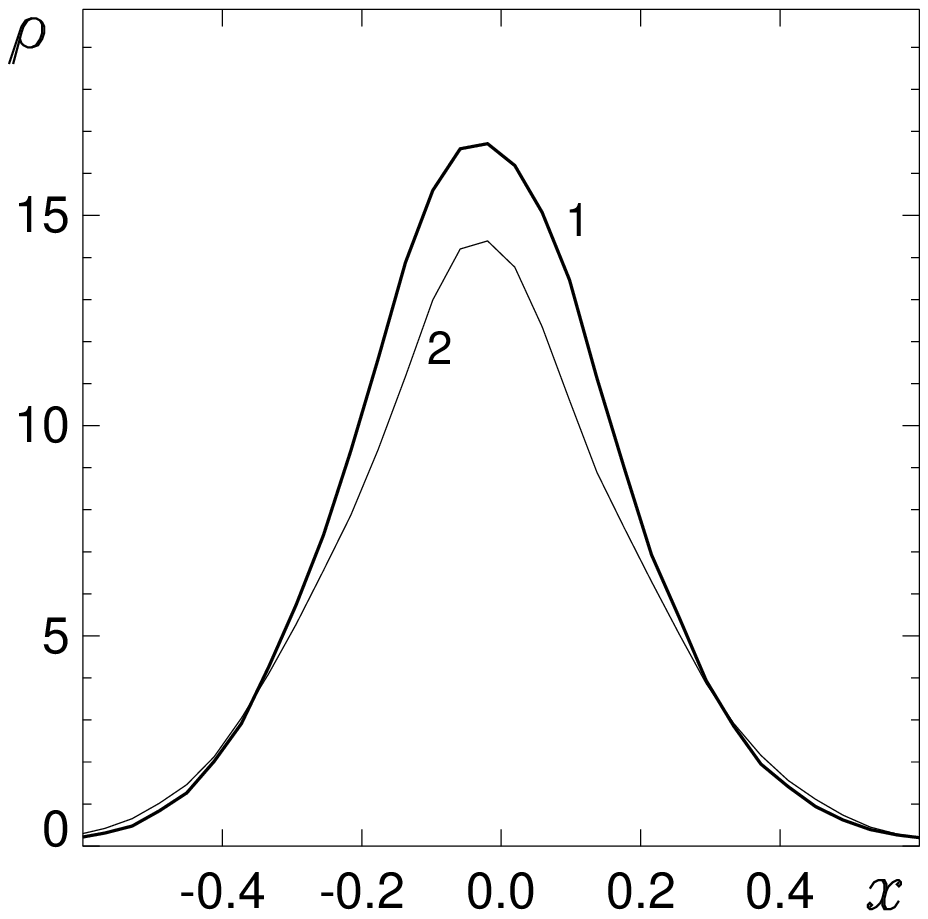,height=5.cm,angle=0} }
\caption{\small {The change in density profile for the galaxy as a
result of the interaction with a shell of fragments dragged in
towards the galaxy center, for the case of $N_f = 20$ (run
$BS_1$). The thicker curve (1) represents the initial density
profile, the thinner (2) gives the final profile.}}
\label{fig:fig15}
\end{figure*}

The use of the smoother initialization described as case II in
Sect.~\ref{sect:case2}
leads to runs where the secular effects associated with
the granularity of the fragment population are more convincingly
identified. The orbital decay of the fragments is illustrated in
Fig.~\ref{fig:fig13},
where we also show for comparison a parallel run where
the shell is treated as collisionless. The interaction which
generates dynamical friction of the galaxy on the fragments is
responsible for the slow, but systematic, development of a
tangential bias in the pressure tensor, as illustrated in Fig.~\ref{fig:fig14}
and for an overall change in the density profile of the galaxy
(see Fig.~\ref{fig:fig15}).

The check against the collisionless run is worth a special
digression. In order to  diagnose the collisionality present in
our system better,  we have performed the following test. We have
studied the correlation between {\it single particle energy} (for
the particles simulating the galaxy) at two different times. If
the simulation is truly collisionless, the single particle energy
$E_p = m v^2/2 + m \Phi(r)$, where $\Phi(r)$ is the mean field
potential, should be strictly conserved. Indeed, by means of a
plot similar to the lower right frame of Fig.~\ref{fig:fig13}, we have checked
that for the run illustrated by the lower left frame of Fig.~\ref{fig:fig13}
more than $90 \%$ of the particles have an energy correlation such
that $|[E_p(t=118~t_{cr})-E_p(t=14~t_{cr})]/E_p(t=14~t_{cr})|<
0.06$. In turn, for run $BS_2$, otherwise similar, but
characterized by the presence of 100 fragments, a similar level of
correlation for single particle energy (at 6 \%) is preserved
within the same time interval only by 50 \% of the particles; the
lower right frame of Fig.~\ref{fig:fig13} thus illustrates the
``damage" associated with dynamical friction.

\subsubsection{Case III}
\label{sect:many_result3}
\begin{figure*}[hbt!]
\centerline{\epsfig{file=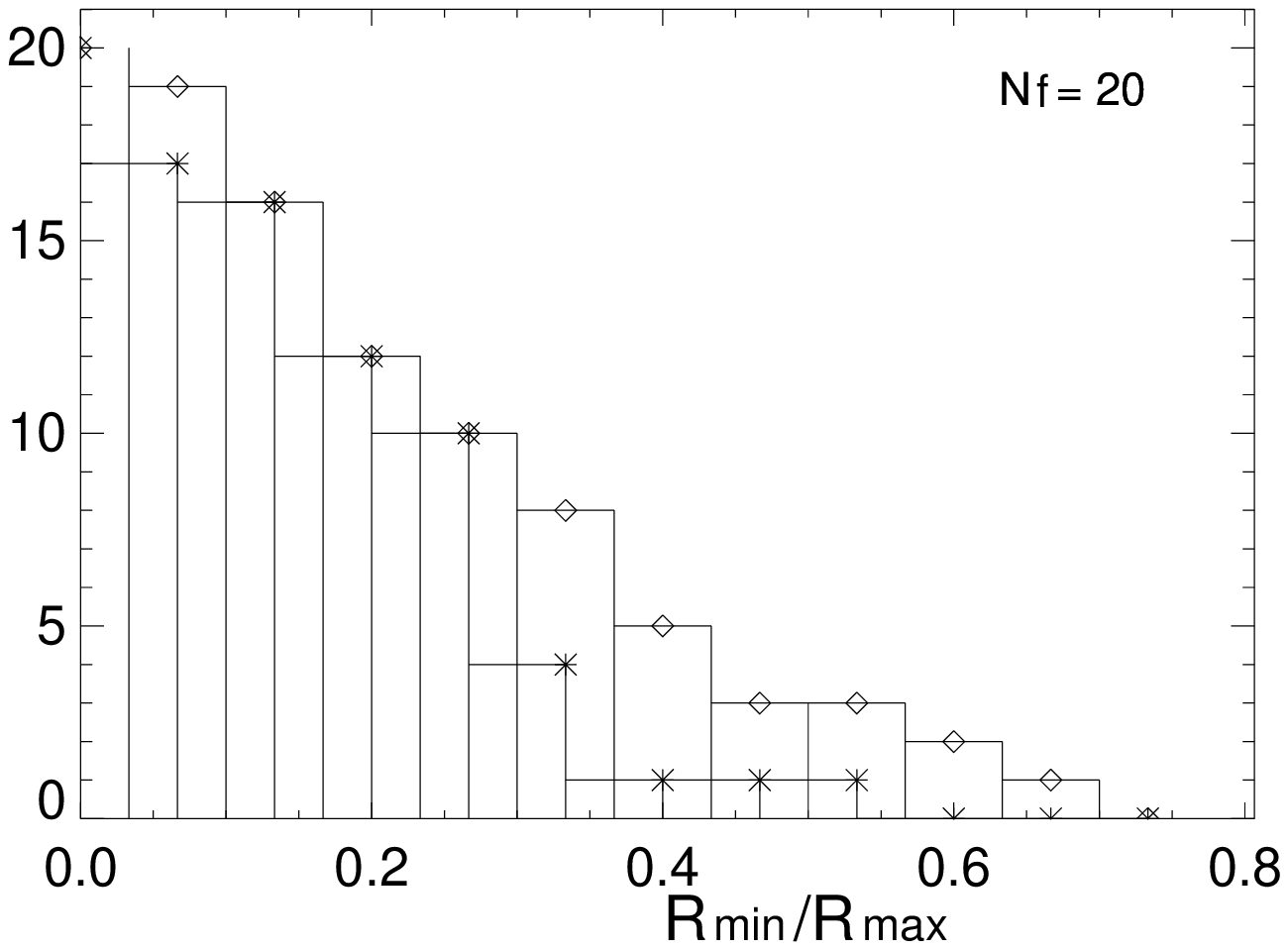,height=4.8cm,angle=0}
\epsfig{file=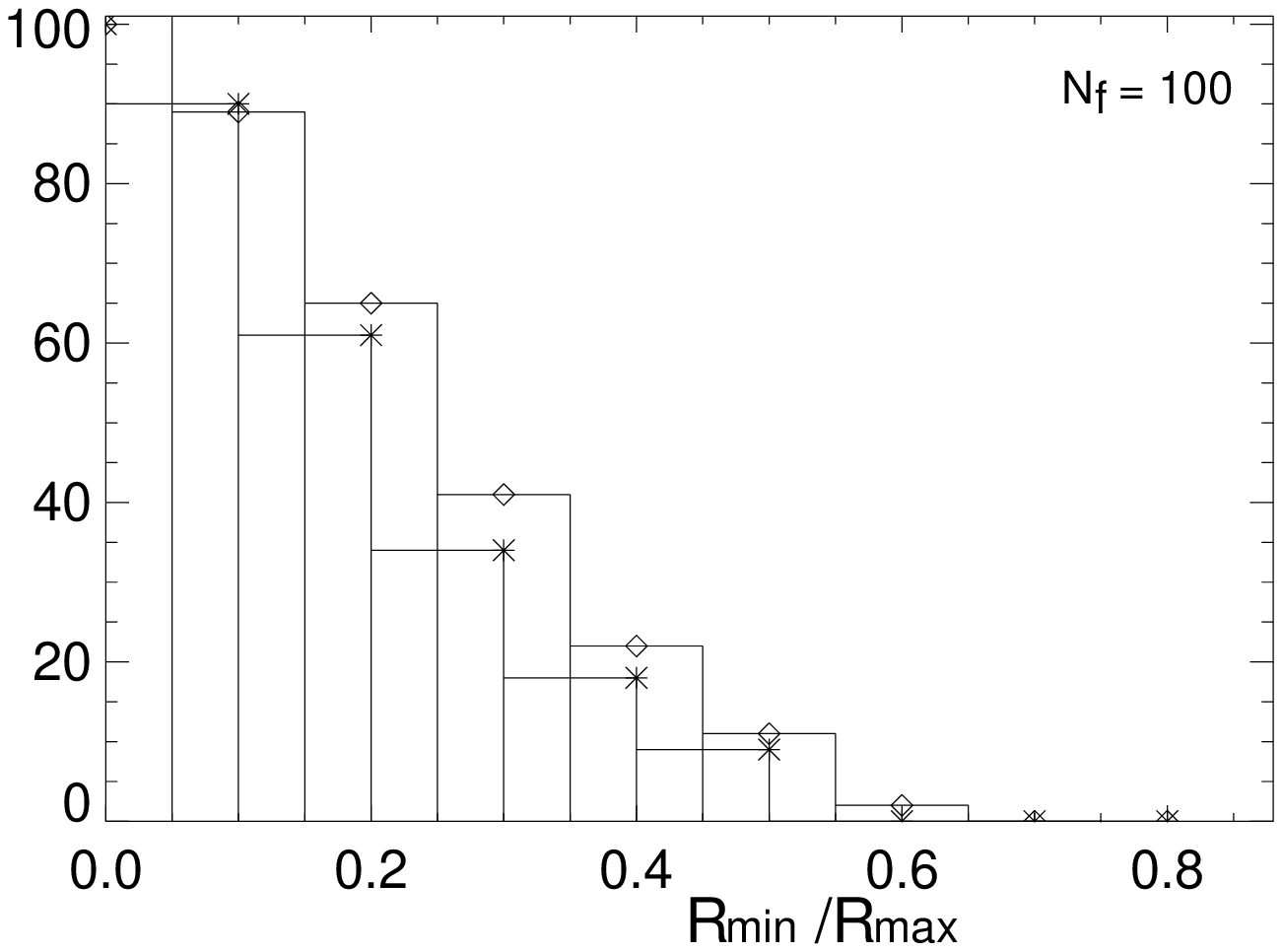,height=4.8cm,angle=0}}
\leftline{
\epsfig{file=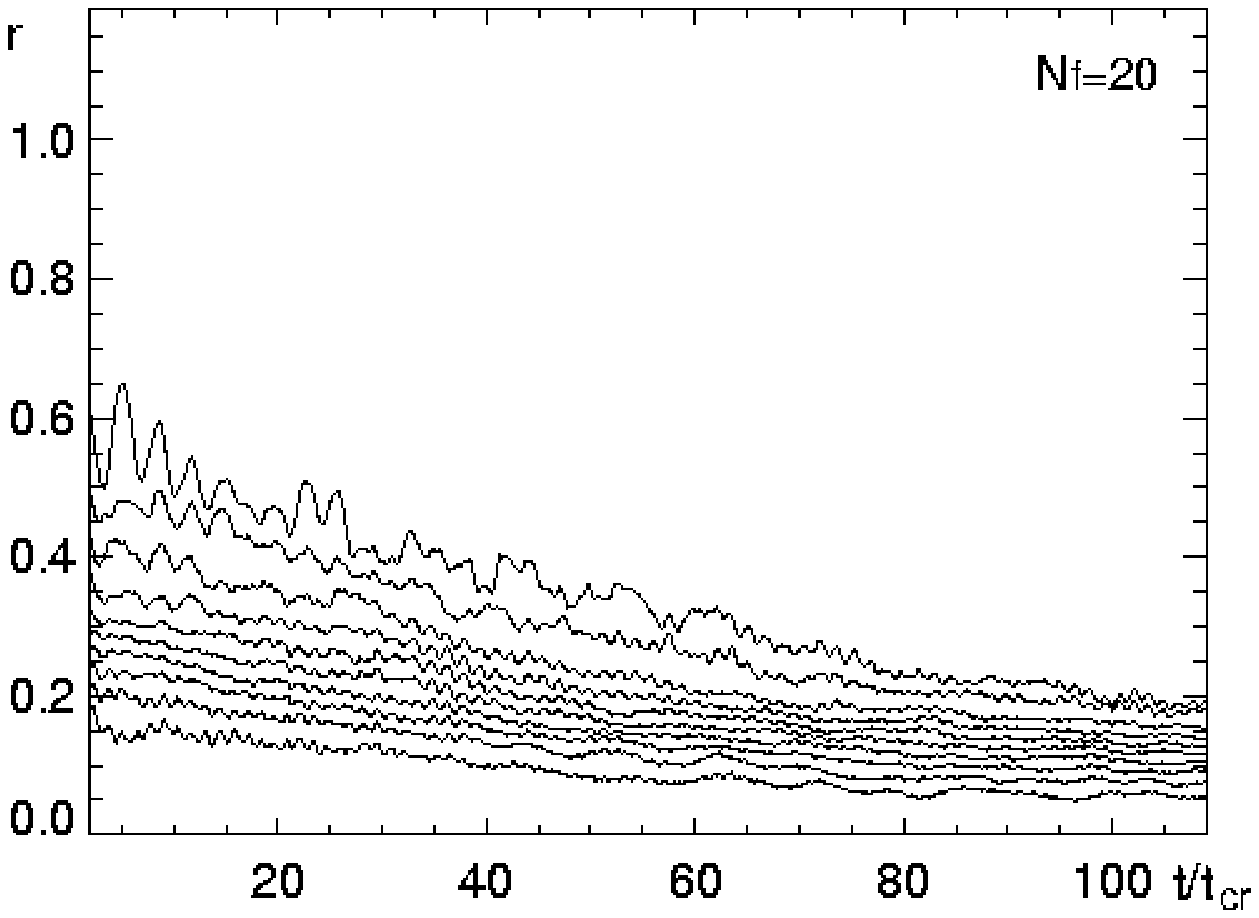,height=50mm,width=65mm,angle=0}
\epsfig{file=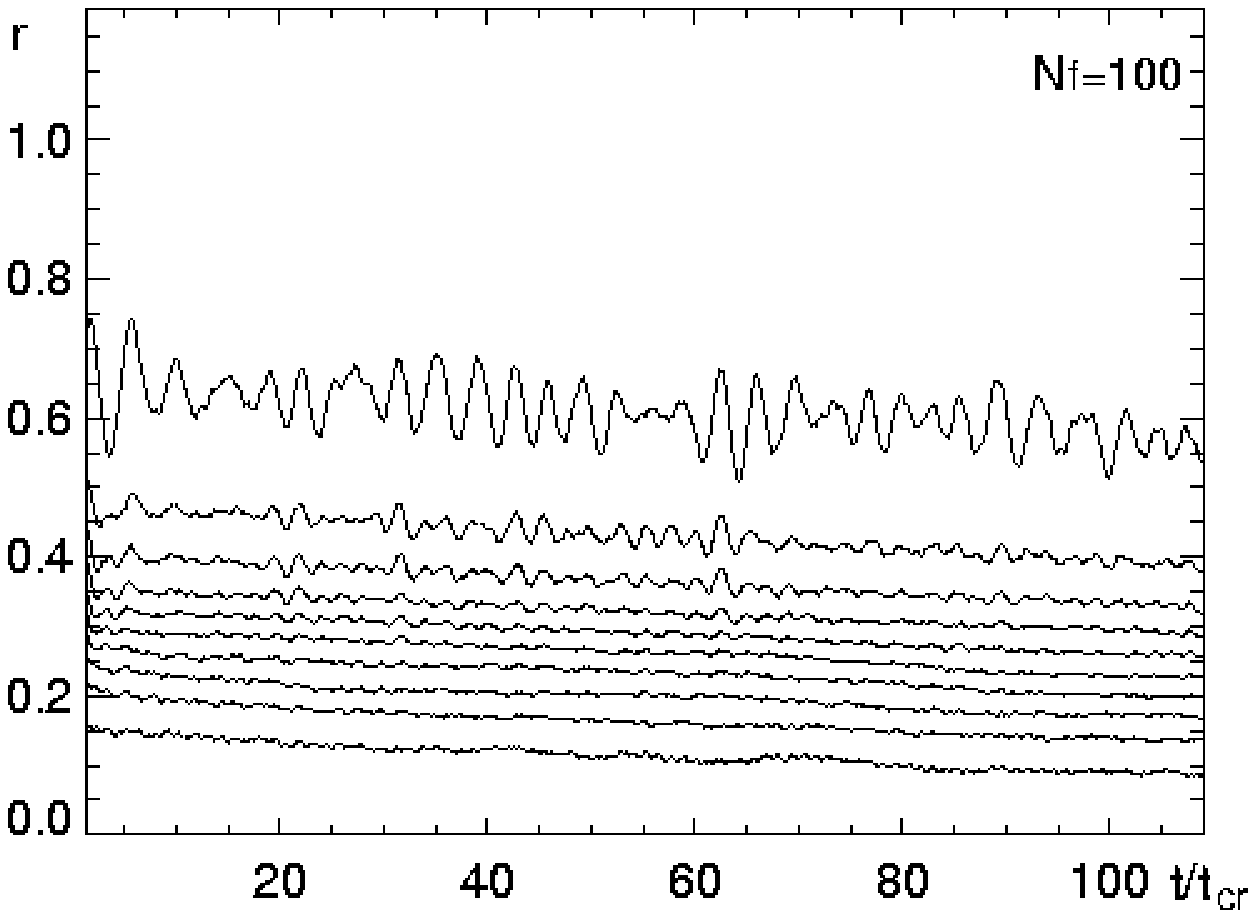,height=50mm,width=65mm,angle=0}}
\caption{\small
The effects of dynamical friction on a shell of fragments created
out of the initial distribution function (case III of Sect.~\ref{sect:model}).
The lower frames show the general decay of orbits of the fragments
for the case $N_f = 20$ (run $BT_3$) and $N_f = 100$ (run $BT_4$);
displayed lines represent 5, 15, 25, 35, 45, 55, 65, 75, 85, and
95 \% of the shell mass. For the two runs, the upper frames show
the corresponding effect of circularization of orbits; here the
histograms record the number of fragments (crosses represent
initial conditions, open circles the conditions at the end of the
run) with orbit aspect ratio above $R_{\rm min}/R_{\rm max}$,
where $R_{\rm min}$ and $R_{\rm max}$ are consecutive minimum and
maximum radii attained by a fragment along its orbit.}
\label{fig:fig16}
\end{figure*}

The even smoother procedure of case III leads to slower orbital
decays, as illustrated in the lower frames of Fig.~\ref{fig:fig16}. For this
type of simulations, the initial distribution of fragment orbits
replicates the isotropic star distribution of the underlying
galaxy. Because of dynamical friction, the fragment orbits are
subject to a slow process of circularization, as illustrated in
the upper frames of Fig.~\ref{fig:fig16}.

\begin{figure*}[hbt!]
\centerline{\epsfig{file=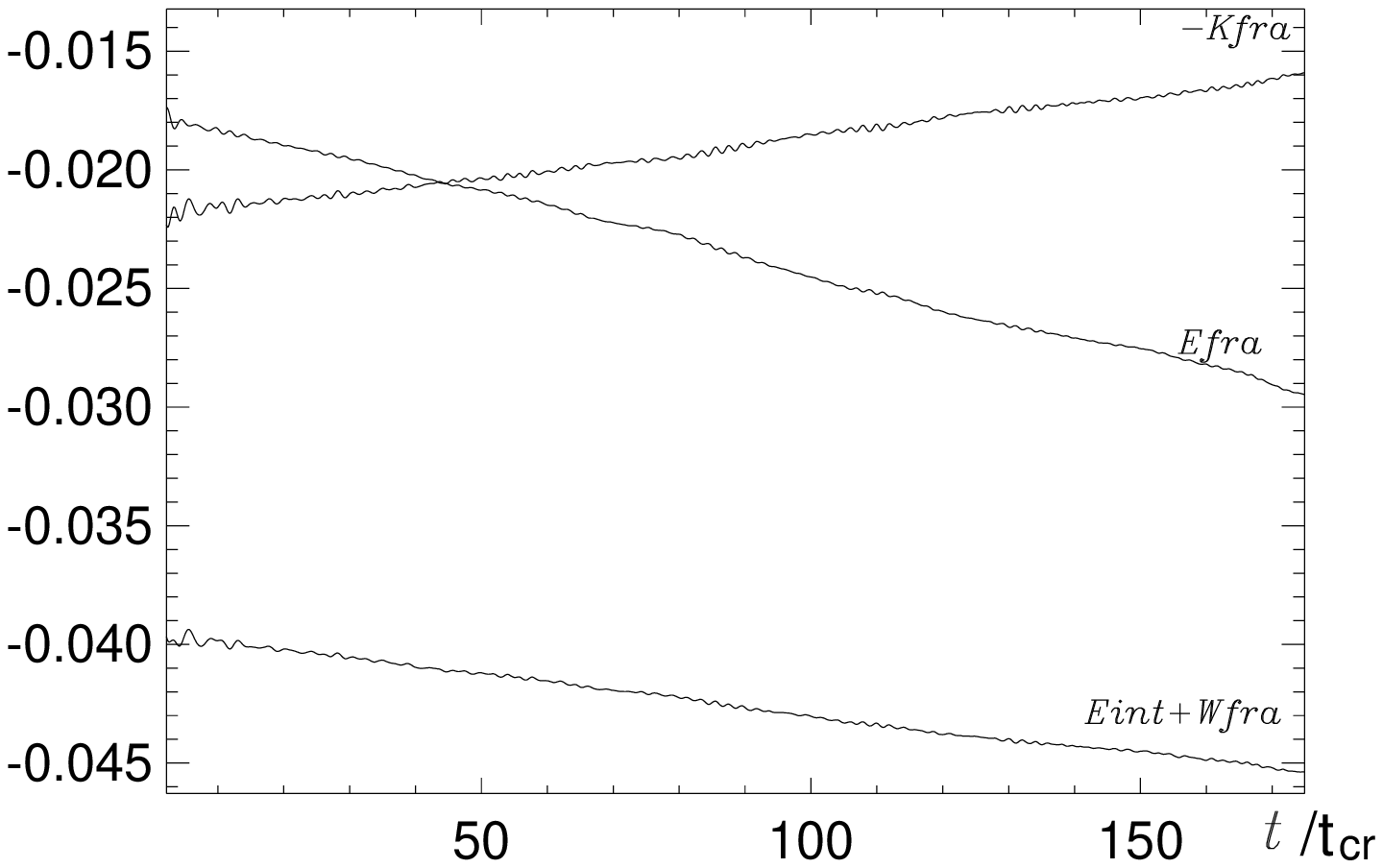,height=4.8cm,angle=0}
\epsfig{file=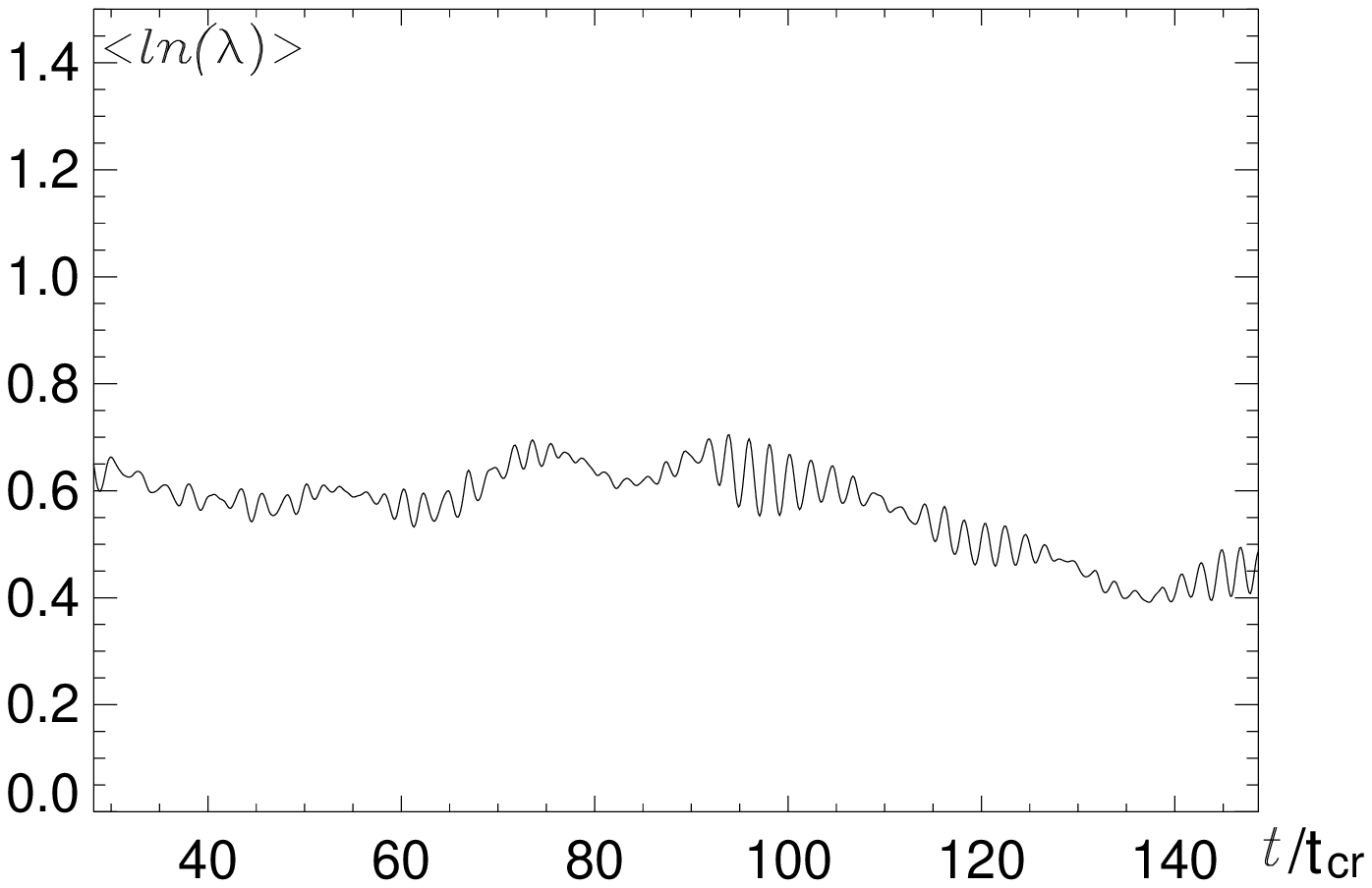,height=4.8cm,angle=0}} \caption{\small The
energies associated with the system of minisatellites (left frame)
and measurement of the Coulomb logarithm $<ln(\lambda)> =
\sum_{j=1}^{j=N_f} ln(\lambda)/N_f$ (right frame) (run $BS_2$).}
\label{fig:fig17}
\end{figure*}

For each of the fragments that are involved in the simulation
described as Case II, we can repeat the analysis that has led to
the measurement of the Coulomb logarithm from the study of the
decay of the orbit of a single satellite.  For example, for run
$BS_2$ we have 100 independent such determinations. In
Fig.~\ref{fig:fig17} we record, for run $BS_2$ a measurement of an
average value of the Coulomb logarithm. We note that the fact that
the value of the Coulomb logarithm thus determined remains of the
same order of magnitude as that obtained in Fig.~\ref{fig:fig1}
proves that the classical expression for the effects of dynamical
friction captures the correct scaling with respect to the mass of
the object subject to friction.

\section{Discussion and conclusions}
\label{sect:conclusion}
In this paper we have set up a numerical laboratory for the study
of dynamical friction in inhomogeneous quasi-spherical galaxies
and made several experiments that have elucidated the physical
mechanisms that participate in the process.

The general character of the simulations presented here is
distinctly different from that adopted in previous studies in this
general research area, because we have addressed a physical
scenario in which the slow collapse induced by dynamical friction
approximately preserves the spherical symmetry present initially.
This configuration is inspired by the astrophysical problem of
studying the evolution of a spherical galaxy in the presence of a
substantial globular cluster system or as a result of the capture
of a large set of mini-satellites dragged in along random
directions.

From the theoretical point of view, the framework adopted here has
an important advantage with respect to the traditional case of the
study of the decay of the orbit of a single satellite, because it
allows us to run smoother simulations that are basically free from
other effects unrelated to dynamical friction, such as those
associated with lack of equilibrium in the initial configuration.

In the past, most of the attention has been focused on the effect
of the host galaxy on the object subject to dynamical friction.
Here, we have been able not only to test the adequacy of the
classical formulae of dynamical friction (derived for ideal
homogeneous models) in the inhomogeneous galaxy environment, but
also to measure the evolution of the density and pressure tensor
profiles in the galaxy induced by the presence of friction
processes on a minority population of heavier particles. We have
made quantitative tests that convince us that we are detecting
genuine effects associated with secular evolution.

\subsection{Plans for future investigations}
\label{sect:plans}
Starting from the encouraging quantitative results obtained in
this paper, there are now two major issues that we would like to
address in investigations planned for the near future, which
appear to be approachable by the use of the numerical laboratory
developed here:

\begin{itemize}

\item In this paper we have limited our study to the presence of
one or two shells of heavy particles, so as to probe the reaction
of the galaxy distribution function in a controlled way and to
test whether the interaction involved can be described as
``local". The first immediate objective that we have in mind is to
devise a set of more realistic simulations, able to represent the
distribution of the globular cluster system in galaxies better.

\item So far we have limited our discussion to the study of the
orbital decay in an $n = 3$ polytropic, isotropic galaxy model.
Now we would like to test systematically the process of slow
collapse induced by dynamical friction as a function of the
concentration properties of the equilibrium model, for the case
of more realistic galaxy models.

\end{itemize}

\begin{acknowledgements}
      Part of this work was supported by the Italian MIUR. We thank Tjeerd van Albada for providing us with a
      copy of his code and for a number of useful conversations.
\end{acknowledgements}

\end{document}